\documentclass[traditabstract]{aa} % 
 \usepackage{graphicx}
\usepackage{color}
\usepackage{savesym}
\savesymbol{rotate}
\usepackage{deluxetable} 
\restoresymbol{DTBL}{rotate}
\usepackage{txfonts} 
\usepackage{mathrsfs}  

\usepackage{natbib}
\usepackage{graphicx}
\bibpunct{(}{)}{;}{a}{}{,} 
\usepackage[english]{babel}
 \usepackage{longtable}
\usepackage{url}
\usepackage{hyperref}
\usepackage[normalem]{ulem}
\usepackage{amstext}

\begin{document} 

\title{The Carnegie Supernova Project II. Observations of the intermediate-luminosity red transient SNhunt120\thanks{This paper includes data gathered with the 6.5 meter Magellan telescopes at Las Campanas Observatory, Chile.}$^{,}$\thanks{Photometry and spectra presented in this paper are available on \href{https://wiserep.weizmann.ac.il/}{WISeREP}.}}

\author{M. D. Stritzinger\inst{1}
\and F. Taddia\inst{1}
\and M. Fraser\inst{2}
\and T.~M. Tauris\inst{3,1}
\and N. B. Suntzeff\inst{4,5}
\and C. Contreras\inst{1,6}
\and S Drybye\inst{1,7}
\and L. Galbany\inst{8}
\and S. Holmbo\inst{1}
\and N. Morrell\inst{6}
\and M. M. Phillips\inst{6}
\and J. L. Prieto\inst{9,10}
\and J. Anais\inst{6}
\and C. Ashall\inst{11}
\and E. Baron\inst{12}
\and C. R. Burns\inst{13}
\and P. Hoeflich\inst{11}
\and E. Y. Hsiao\inst{11}
\and E. Karamehmetoglu\inst{1}
\and T. J. Moriya\inst{14,15}
\and M. T. Botticella\inst{16}
\and A. Campillay\inst{6}
\and S. Castell\'{o}n\inst{6}
\and C. Gonz\'{a}lez\inst{6}
\and M. L. Pumo\inst{17,18,19}
\and S. Torres-Robledo\inst{20,6}
}

\institute{Department of Physics and Astronomy, Aarhus University, 
Ny Munkegade 120, DK-8000 Aarhus C, Denmark 
\and 
School of Physics, O'Brien Centre for Science North, University College Dublin, Belfield, Dublin 4, Ireland
\and
Aarhus Institute of Advanced Studies (AIAS), Aarhus University, H{\o}egh-Guldbergs Gade 6B, 8000 Aarhus C, Denmark
\and
The George P. and Cynthia Woods
Mitchell Institute for Fundamental Physics and Astronomy, Texas A\&M University, College Station, TX 877843
\and
Department of Physics and Astronomy, Texas A\&M University, College Station, TX 77843
\and
Las Campanas Observatory, Carnegie Observatories, Casilla 601, La Serena, Chile %0000-0003-2734-0796
\and
Nordic Optical Telescope, Apartado 474, E-38700 Santa Cruz de La Palma, Spain
\and
Departamento de F\'isica Te\'orica y del Cosmos, Universidad de Granada, E-18071 Granada, Spain
\and
N\'ucleo de Astronom\'ia de la Facultad de Ingenier\'ia y Ciencias, Universidad Diego Portales, Av. Ej\'ercito 441 Santiago, Chile
\and
Millennium Institute of Astrophysics, Santiago, Chile
\and
Department of Physics, Florida State University, Tallahassee, FL 32306, USA
\and
Homer L. Dodge Department of Physics and Astronomy, University of Oklahoma, 440 W. Brooks, Rm 100, Norman, OK 73019-2061, USA
\and
Observatories of the Carnegie Institution for Science, 813 Santa Barbara St, Pasadena, CA, 91101, USA
\and
National Astronomical Observatory of Japan, National Institutes of Natural Sciences, 2-21-1 Osawa, Mitaka, Tokyo 181-8588, Japan
\and
School of Physics and Astronomy, Faculty of Science, Monash University, Clayton, VIC 3800, Australia
\and
INAF- Osservatorio Astronomico di Capodimonte, Salita Moiarello 16,
Napoli, Italy
\and
Universit\`a degli studi di Catania, Dip.~di Fisica e Astronomia ``E. Majorana'', Via Santa Sofia 64, I-95123, Catania, Italy
\and
INAF - Osservatorio Astronomico di Padova, Vicolo dell'Osservatorio 5, 35122, Padova, Italy
\and
Laboratori Nazionali del Sud-INFN, Via Santa Sofia 62, I-95123, Catania, Italy
\and
SOAR Telescope, La Serena 1700000, Chile} 

  \abstract
   {We present  multiwavelength observations of two gap transients
   that were followed by the Carnegie Supernova Project-II. The observations are supplemented  with data  obtained by a number of different programs. Here in the first of two papers, we focus on  the intermediate-luminosity red transient (ILRT) designated SNhunt120, while in a  companion paper we examine the luminous red novae AT~2014ej.
   Our data set for SNhunt120 consists of an early optical discovery, estimated to be within three days after outburst,  the subsequent optical and near-infrared broadband followup extending over a period of $\text{about two}$  months, two  visual and two near-infrared wavelength spectra, and \textit{Spitzer} Space Telescope observations extending from early ($+$28~d) to late ($+$1155~d) phases.   SNhunt120  resembles other ILRTs such as NGC~300-2008-OT and SN~2008S, and like these other ILRTs, SNhunt120 exhibits prevalent mid-infrared emission at both early and late phases. From  the comparison of SNhunt120 and other ILRTs to electron-capture supernova simulations, we find  that the current models underestimate the explosion kinetic energy and thereby produce synthetic light curves that overestimate the luminosity. Finally, examination of pre-outburst \textit{Hubble} Space Telescope images yields no  progenitor detection.}
   \keywords{star: mass loss, circumstellar matter - supernova: individual SNhunt120 (PSN J14535395+0334049, LSQ12brd)}
\authorrunning{Stritzinger, Taddia, Fraser, et al.}     
\titlerunning{CSP-II observations of the ILRT SNhunt120}    
 \date{Received 09 March 2020; accepted }

\maketitle

\section{Introduction}
\label{introduction}

The proliferation of sky surveys over the past decade has driven a dramatic increase in the discovery of transients in the  ``luminosity gap'' between the brightest novae ($M_V \lesssim -10$ mag) and the faintest classical core-collapse supernovae (SN; peak $M_V \sim -16$ mag).  
Because they resemble some type~IIn supernova, but have a much fainter peak luminosity, these objects have been referred to in the past years as SN imposters  \citep[e.g.,][]{1990MNRAS.244..269S,1997ARA&A..35..309F,2000PASP..112.1532V}. However, with the discovery of a low-luminosity subclass of type~IIP   SNe \citep[e.g.,][]{2004MNRAS.347...74P} and an ever-increasing diversity of observed phenomena, the SN imposter nomenclature has now become  obsolete, and today, SN imposters are typically associated with luminous blue variables (LBVs; \citealt[e.g.][]{2011MNRAS.415..773S}). 
Gap transients have also been referred to as intermediate-luminosity optical transients (ILOTs), but we choose not to use this term to avoid confusion with intermediate-luminosity red transients (ILRTs; \citealt[e.g.]{bond_ngc300ot}), as well as with luminous red novae (LRNe; \citealt[e.g.][]{2017ApJ...834..107B}).  
The early spectra of gap transients are often dominated by Balmer emission lines, typically with full width at half-maximum (FWHM) emission line velocities ranging  from $\gtrsim$ 200-300~km~s$^{-1}$ to $\gtrsim$ 1000~km~s$^{-1}$.
This can be taken as evidence for the presence of a preexisting circumstellar medium (CSM) around the progenitor of these transients.

Evidence now indicates that different observational subtypes of gap transients are linked to different progenitors \citep[see][for concise reviews]{2009aaxo.conf..312K,2012ApJ...758..142K,kashi2016,2019NatAs...3..676P}.
In Fig.~\ref{Fig:Mr_vs_time} the regions within the peak absolute magnitude versus decay-time parameter space that harbor the different flavors of gap transients are indicated, including  LBVs, ILRTs, and LRNe. The figure also includes regions of other known transient populations.
LBVs are stars that undergo giant eruptions and display erratic variability that in some cases may occur immediately prior to a star exploding as a core-collapse supernova. 
ILRTs are commonly linked to super asymptotic giant branch (S-AGB) stars \citep{2008ApJ...681L...9P,thompson09,2009MNRAS.398.1041B,2009ApJ...705.1425P,2011ApJ...741...37K,2016MNRAS.460.1645A,2017PASA...34...56D} that undergo a weak terminal explosion induced by an electron-capture collapse \citep{1980PASJ...32..303M,1984ApJ...277..791N,1987ApJ...318..307M,1993ApJ...414L.105H,2006A&A...450..345K,2008ApJ...675..614P}. 
Other models have also been presented to account for ILRTs. 
For example, it has been proposed that ILRTs are associated with moderately massive ($\sim$ 20~$M_{\sun}$) stars  embedded in a dusty environment that experience outbursts driven by super-Eddington winds \citep[e.g.,][]{2009ApJ...697L..49S,2011ApJ...743..118H}.  Alternatively, \citet{kashi2010} presented a model for ILRTs  being  powered by the gravitational energy release following the accretion of matter from an AGB star  onto a main-sequence or more compact companion.
    
Finally, LRNe may be linked  to massive contact binary systems \citep[e.g.,][]{2017ApJ...834..107B}   that undergo a common-envelope ejection while the two stars  merge \citep{2016MNRAS.458..950S,2017MNRAS.470.2339L,2018MNRAS.473.3765M}. 
 This scenario is  the massive star version
  of common-envelope ejections associated with less massive stars producing red novae with double-peak light curves. 
 Today it remains a matter of open debate whether the merger process driving LRNe  culminates in the coalesce of a single star, a weak SN explosion, or even both, depending on the masses and pre-SN mass-loss histories  \citep[see][for a detailed discussion]{2019A&A...630A..75P}.

Only a relatively small number of gap transients have been studied in the literature, therefore we present in two papers a detailed study of two  gap  transients observed  by the \textit{Carnegie Supernova Project} (CSP-II; \citealt{2019PASP..131a4001P}).
These objects, designated 
SNhunt120 and AT~2014ej,  are both plotted in Fig.~\ref{Fig:Mr_vs_time} along with  the handful of other well-observed LRNe and ILRTs from the literature. 
Inspection of the figure reveals that SNhunt120 is located  within the ILRT gap transient region, and AT~2014ej is located within the LRNe gap transient region. 

Because so many data have been obtained for the two objects and because our analysis is extended,  we  elected  to present our study of SNhunt120 and AT~2014ej in separate papers. Here in Paper~1 we focus on   SNhunt120, while in the companion paper  (Stritzinger et al. 2020c; hereafter Paper~2)   we focus on AT~2014ej. 
In addition to CSP-II data presented here in Paper~1, we also   make use of  early data obtained by  the La Silla-Quest  (LSQ) low-redshift supernova survey  \citep{2013PASP..125..683B} to estimate the outburst time of SNhunt120. The analysis we present below also makes use of optical photometry previously presented by \citet{2018RNAAS...2c.176B},  two-channel
\textit{Spitzer} Space Telescope images  spread over 1000 days, and  pre-explosion \textit{Hubble} Space Telescope (HST) images of the host galaxy NGC~5775.

\section{SNhunt120}  
\subsection{Discovery, distance, and reddening}
\label{SNhunt120details} 

SNhunt120, also known as  PSN J1453595$+$0334049 and LSQ12brd, was discovered by the Catalina Sky Survey (CSS) on 27.49 March 2012 UT \citep{2012ATel.4004....1H} in NGC~5775 with an  apparent $V$-band magnitude $m_V = 18.7$ mag. The transient was not detected in CSS images taken on 17.39 March 2012 UT down to a limiting magnitude $m_V > 19.5$.
We have also recovered the transient in  an  LSQ image obtained on 27.30 March 2012 UT with an apparent $gr$-band magnitude of 18.95.

A spectrum obtained on 28.16 March 2012 UT  with the 2.5m du Pont ($+$WFCCD) telescope  at Las Campanas Observatory (LCO), reported by \citet{2012ATel.4004....1H}, exhibits Balmer emission lines and both allowed and forbidden \ion{Ca}{ii} features. Furthermore, the H$\beta$ $\lambda$4861 and H$\alpha$ $\lambda$6563  Balmer lines are characterized by FWHM velocities of  $v_{FWHM} \sim 650$ km~s$^{-1}$. 
These  characteristics, along with the fact that the transient reached an absolute $V$-band magnitude $M_V \approx -13.9$ mag, led \citet{2012ATel.4004....1H}  to identify it as a possible gap transient similar to SN~2008S \citep{2009MNRAS.398.1041B} and NGC~300-2008-OT \citep{bond_ngc300ot,2011ApJ...743..118H}. 

\citet{2012ATel.4009....1B} reported on the possible detection of a source at the position of this gap transient in  a pre-explosion archival HST (+ ACS/F625W) image of NGC~5775, measuring an apparent magnitude of 25.5-26, and provided corresponding absolute magnitudes of $\sim$ $-6.1$ to  $-6.6$, assuming a distance of 31~Mpc. This is larger than the value we adopt.  For our distance value (see below), the absolute magnitude would be between $-$4.7 and $-$5.2 mag. 

With J2000 coordinates of $\alpha=14^{\rm h}53^{\rm m}53\fs95$ and $\delta=+03^{\circ}34\arcmin04\fs9$, SNhunt120 was located 55$\arcsec$ west and 85$\arcsec$ north from the core of its host NGC~5775 (see Fig.~\ref{fig:FC1} for a finding chart).  
NGC~5775 is an edge-on SBc spiral galaxy and is a member of the Virgo cluster. 
The NASA/IPAC Extragalactic Database (NED)\footnote{\url{http://ned.ipac.caltech.edu}} lists its heliocentric redshift as $z = 0.005607$ and also lists over half a dozen different Tully-Fisher distances from the literature, ranging from 15.9 to 27.0 Mpc (distance moduli from 31.0$\pm$0.40 mag to 32.15$\pm$0.40 mag).
However, as NGC~5775 is a member of the Virgo cluster, we opt to use the more precise distance modulus to 
Virgo of $\mu$~$=$~31.03$\pm$0.14 mag, which is based on a Cepheid calibration of the Tully-Fisher relation as presented by \citet{tully2016}.

 In Appendix~\ref{appendixA} we compute the host redshift and metallicity using a spectrum of NGC~5775 obtained by the Sloan Digital Sky Survey, while Appendix~\ref{appendixB} contains details on the Galactic reddening in the direction of SNhunt120 and our estimate of its host-galaxy reddening. In short, in the work presented below, we adopted a total visual extinction of $A^{tot}_{V} = 0.76\pm0.25$ mag.

\subsection{Observations}
\label{SNhunt120observations}

The CSP-II obtained 22 epochs of  $ugriBV$-band images of SNhunt120 with the Swope (+ CCD camera) telescope located at LCO  extending between $+$2 to $+$46 days (d) past first detection. Two epochs of $YJH$-band NIR photometry were obtained with the du Pont (+ RetroCam; \citealt{hamuy2006}) telescope on $+$16~d and $+$37~d, while two epochs of $Y$-band photometry were obtained nearly a day apart on $+$153~d and $+$154~d. A single $J$-band observation was also obtained a day later. The raw data were reduced using standard CSP-I pipelines as described in detail by \citet{2017AJ....154..211K} and references therein. Prior to computing photometry, deep host-galaxy template images were subtracted from all of the optical science images, but not for the NIR science images because we lack template images.

Photometry of SNhunt120 was computed relative to a local sequence of stars calibrated with respect to optical and NIR standard star fields observed over a minimum of three photometric nights. The optical photometry of these stars in the ``standard'' system is presented in Table~1, while  NIR photometry of a smaller set of stars in the standard system is given in Table~\ref{tab:SNhunt120_nir_locseq}.
Finally, our  optical and NIR photometry of SNhunt120 in the CSP-II `natural'  photometric systems are listed in  Table~\ref{tab:SNhunt120_optphot} and Table~\ref{tab:SNhunt-nirphot}, respectively. 

The La Silla Quest (LSQ) survey observed  NGC~5775 multiple times over a period of several years, and we are able to identify SNhunt120 in 17 epochs of LSQ $gr$-band imaging. 
Prior to computing photometry of SNhunt120 from these images, a deep galaxy template image was constructed from a series of 15 LSQ images obtained prior to discovery. After subtracting the template image from the science images, point spread function (PSF) fitting photometry of SNhunt120 was computed relative to the $V$-band magnitude of a single bright star in the field of NGC~5775, which is calibrated by the CSP-II data and designated with the ID of 3 (see below). 
Following \citet[][see their Fig. 23]{contreras18}, PSF photometry  was computed from the LSQ $gr$-band images and calibrated relative to the Swope natural $V$-band photometric system. The  LSQ photometry  is  listed in Table~\ref{tab:SNhunt120_LSQ_phot}.

A search of the Spitzer data archive reveals images were obtained of SNhunt120 over several epochs with the 
3.6~$\mu$m and 4.5~$\mu$m  channels.
Specifically, the transient was detected during two early epochs ($+$28~d and $+$39~d), and also at late times on $+$1155~d.
Fully reduced stacked images were downloaded from the Spitzer archive. PSF photometry was computed with a pipeline developed in \texttt{Matlab,} and the results are listed in Table~\ref{tab:spitzer}. These fluxes are used below to construct spectral energy distributions (SEDs) for SNhunt120 extending from the atmospheric cutoff out to 4.5~$\mu$m.

Turning to spectroscopy of SNhunt120, two epochs of visual wavelength observations are presented. This includes the classification spectrum and a followup spectrum taken by the CSP-II. These are complemented with  two epochs of NIR spectroscopy also obtained by the CSP-II, which to our knowledge represent the first NIR spectra published of an ILRT.
A summary of these observations is provided in the log of spectroscopic observations in Table~\ref{tab:specobs}. The visual wavelength spectra were obtained with the du Pont ($+$ WFCCD) telescope on $+$0.9~d and $+$35.7~d post discovery, while the NIR spectra were taken with the  Magellan Baade ($+$ FIRE) telescope on $+$11.9~d and $+$40.7~d. The optical data were reduced following standard methods as described in \citet{hamuy2006} and \citet{2017AJ....154..211K} and the NIR reductions following the prescription of \citet{2019PASP..131a4002H}.

\section{Results}
\subsection{Photometry, explosion epoch, intrinsic broadband colors, and spectral energy distributions}
\label{SNhunt120photometry}

First we examine the discovery and early followup photometry of SNhunt120 obtained by CSP-II, LSQ, CSS, and \citet{2018RNAAS...2c.176B} and plotted in Fig.~\ref{LSQearly}. Overplotted on the  photometry is the best-fit power-law function fit to the first six epochs of observations, characterized by a power-law index of $\alpha = 0.57\pm0.20$. Extrapolating to zero flux provides an  explosion epoch on JD--$2456012.58\pm0.95$, which is $1.25$ days prior to the first LSQ detection. 

Now we proceed to examine the full set of broadband photometry obtained for SNhunt120. Optical ($ugriBV$) and NIR ($YJH$) light curves of SNhunt120 are plotted in  Fig.~\ref{Fig:SNhunt120photometry}. 
The optical light curves of SNhunt120 exhibit a $\sim$0.5 mag rise to maximum over the first week of evolution covered by our observations.
Upon reaching maximum, the optical light curves declined by $\approx$ 1.0 mag over the period of a month. 

At maximum, SNhunt120 reached an apparent magnitude of $m_V =  17.78\pm0.01$ mag, which when corrected for the adopted extinction  using a standard reddening law and distance corresponds to an absolute  magnitude of $M_V =  -14.03\pm$0.29 mag. Here the accompanying uncertainty is computed by adding in quadrature the uncertainties on $A^{host}_{V}$ and on the adopted distance. 
A summary of peak apparent and absolute peak magnitudes for each of the optical bands is listed in Table~\ref{tab:SNhunt120peaks}. The peaks were obtained by fitting low-order polynomials to the light curves, and these polynomials are plotted in Fig.~\ref{Fig:SNhunt120photometry}.
Because the reddening estimate is uncertain, peak magnitudes are listed with and without host-reddening corrections.
Inspection of the values listed in Table~\ref{tab:SNhunt120peaks} indicates that the optical light curves all peaked within a day of each other and that the various peak absolute magnitudes exhibit $\sim 0.7$ mag in scatter, with the $u$ band being the faintest (i.e., $M_u = -13.43\pm0.29$ mag) and the $r$ band the brightest (i.e., $M_r = -14.12\pm0.29$ mag).

We plot in Fig.~\ref{Fig:colorSNhunt120} the intrinsic $(B-V)_0$, $(V-r)_0$, and $(r-i)_0$ broadband colors of SNhunt120. Also shown are the intrinsic broadband colors of the well-observed ILRTs M85-2006-OT1 \citep{m85_kulkarni,m85_pastorello}, SN~2008S \citep{2009MNRAS.398.1041B}, NGC~300-2008-OT1 \citep{bond_ngc300ot},   PTF10fqs \citep{kasliwal11}, AT~2017be \citep{at2017be}, and AT~2019abn \citep{jencson2019,at2019abn}.
The photometry of SNhunt120, M85-OT2006-1, SN~2008S, NGC~300-2008-OT1, PTF10fqs, AT~2017be,  and AT~2019abn
were corrected for reddening by adopting the $E(B-V)_{tot}$ values listed in Table~\ref{tab:comp_obj_snhunt120}.
%0.24 mag, 
%0.14 mag,
%0.69 mag, 
%0.40 mag,
%0.039 mag, 
%0.09 mag, 
%0.30 mag, and
%0.85 mag
%respectively, and taken from references listed in the caption of Fig.~\ref{Fig:colorSNhunt120}.
The host extinction of SNhunt120, as discussed in Appendix~\ref{appendixB}, was assumed to be $A^{host}_V = 0.64$ mag in order to match the color curves of the bluest objects of the comparison sample. 
%and  placed on the absolute magnitude scale adopting distance moduli of 
%32.34 mag, 
%31.34 mag
%28.78 mag, 
%26.37 mag,
%31.15 mag, 
%29.47 mag, and
%31.64 mag. 
Over the first days of evolution, while SNhunt120 increased in brightness, its optical colors evolved toward the blue, and this behavior is most apparent in SNhunt120 (and AT~2017be, but with larger error) among the other ILRTs, whose multiband observations started later than those for SNhunt120 (for NGC~300-2008-OT1 about 20 days after first detection, see \citealp{bond_ngc300ot}). Upon reaching maximum, the colors of SNhunt120 transition and evolve to the red as the object continued to expand and cool. The overall evolution of the ILRTs are similar, but PTF10fqs and M85-2006-OT1 exhibit redder  $(r-i)_0$ and/or $(R-I)_0$ colors than the other objects, and the $(B-V)_0$ of AT~2017be is redder than SNhunt120 and the other objects.

We now proceed to use the broadband photometry of SNhunt120 to construct
SEDs and then fit them with blackbody (BB) functions, in order to estimate the BB luminosity ($L^{hot}_{BB}$), a BB-radius ($R^{hot}_{BB}$) and BB-temperature ($T_{BB}^{hot}$) evolution over time.
To do so, the broadband magnitudes  in the Swope natural system were transformed to AB magnitudes \citep{2017AJ....154..211K}, corrected for extinction, and then converted into monochromatic fluxes at the effective wavelengths of each passband. The resulting SEDs and their corresponding best-fit BB functions %(red lines)\LEt{details from the figure captions should not be repeated in the main text; please check this throughout (this also includes "black diamonds" and similar. I will be editing your companion paper next, and this also applies there, of course. } 
are plotted in the left panel of Fig.~\ref{Fig:BB1}, while the right panels display the temporal evolution (from top to bottom) of $L_{BB}^{hot}$, $T_{BB}^{hot}$, and $R_{BB}^{hot}$. To complement our analysis, SEDs are also plotted (dashed lines) based on photometry published by \citet{2018RNAAS...2c.176B}. To facilitate comparison, the  \citeauthor{2018RNAAS...2c.176B} data were shifted to match the CSP-II photometry through the use of a single additive constant for each bandpass.

Inspection of the BB parameter profiles reveals that the $T_{BB}^{hot}$ increases as the light curve rises to  peak,  with $T_{BB}^{hot}$ reaching a maximum value of $\sim$8700~K. Subsequently, as the luminosity declines, $T_{BB}^{hot}$ also declines, reaching   $\lesssim 6000$~K at $+$50~d, whereupon it follows a plateau.  
At early times, the BB radius is about 2.0$\times$10$^{14}$~cm and declines until $+$30~d, whereupon it increases over a three-week period before settling onto a plateau phase.
The comparison between the bolometric quantities derived from the CSP-II data (red diamonds) and those from the \citeauthor{2018RNAAS...2c.176B} data (black empty diamonds), reveal the effect of including $u$ band in the SEDs, which is only available in the CSP-II data. The luminosity is effectively identical, but the temperature is slightly overestimated by $\sim$500--1000~K when the $u$ band is excluded. 

Farther redward, beyond the optical and NIR wavelengths, we also consider the two-channel Spitzer imaging shown in Fig~.\ref{fig:spitzer} (top panel). 
Through linear interpolation of the corresponding photometry, we obtain a measure of the mid-infrared (MIR) brightness of SNhunt120 on $+$36.8~d, when photometry is available in all bands. 
Our early-time extended SED is shown in Fig.~\ref{fig:spitzer} (middle panel).
Inspection of the SED indicates an MIR excess above the flux reproduced by a single BB fit (blue dashed line) characterized by the temperature $T_{BB}^{hot}$  shown in Fig.~\ref{Fig:BB1}. This MIR excess can be reproduced by another warm BB component. However, its temperature is characterized by a large uncertainty because the flux of this warm component is far lower than that of the hot component.
Finally, in the bottom panel of Fig.~\ref{fig:spitzer} we fit the late-phase $+$1155~d Spitzer photometry  with a single BB component characterized by $T_{BB}^{warm} = 577$~K.
We note that there may be legitimate concerns in estimating  $T^{warm}_{BB}$ and $R^{warm}_{BB}$ from BB fits if the underlying SED contained within the wavelength intervals sampled by the Spitzer flux points includes prominent emission or absorption features. However, these concerns are somewhat alleviated when we consider that MIR spectroscopic observations of the ILRT NGC~300-2008-OT published by  \citet{ohsawa2010} revealed an  MIR SED consistent with a smooth continuum lacking prominent emission or absorption features.

\subsection{Spectroscopy}
\label{SNhunt120spectroscopy}

The two epochs of visual wavelength and NIR spectroscopy of SNhunt120 are plotted in  Fig.~\ref{Fig:SNhunt120-spectra}, and their flux calibration was scaled to be consistent with the observed broadband photometry. 
The optical spectra consist of a continuum with superimposed Balmer emission lines (H${\alpha}$, $H{\beta}$, and $H{\gamma}$) and the $[\ion{Ca}{ii}]$ $\lambda\lambda$7291, 7324 doublet. 
In addition, the spectra  exhibit absorption features of $\ion{Na}{i}~D$, $\ion{Ca}{ii}$ H\&K, and the $\ion{Ca}{ii}$ NIR triplet. 
The NIR spectra are to our knowledge the first of an ILRT  published to date. These spectra exhibit a handful of prominent features, including emission lines  arising from the Paschen series: P$\delta$, P$\gamma$, and P$\beta$.
In the second NIR spectrum (+41~d), a narrow emission feature emerges just blueward of the P$\gamma$ feature, which is likely   \ion{He}{i}~$\lambda$10830. 
None of the features exhibited in these spectra show convincing evidence of P-Cygni absorption profiles. 
The BB fit to the spectra in the NIR shows a possible excess above 2 $\mu$m, compatible with the excess observed in Fig.~\ref{fig:spitzer}, where we also make use of MIR photometry.

To determine the velocity associated with the gas producing the most prominent emission features, we fit the emission  profiles with a Lorentzian function, yielding  a measurement of the FWHM velocity (hereafter $v_{FWHM}$). We corrected the measured width of the lines by the spectral resolution to infer the value of $v_{FWHM}$ (see Sect.~4.3 in  \citealp{at2017be}). Some of these velocities are comparable to the instrumental resolution ($\sim$400 km~s$^{-1}$ in the optical and $\sim$600--700 km~s$^{-1}$ in the NIR).  

Figure ~\ref{fig:lineprofilesLSQ12brd} shows the emission features  of H$\beta$, H$\alpha$, the $[\ion{Ca}{ii}]$ doublets, the $\ion{Ca}{ii}$ NIR triplet, \ion{He}{i}~$\lambda$10830, P$\gamma$ and P$\beta$. Each panel contains the data of one of these features (black lines) in velocity space, along with their best-fit Lorentzian profile (red lines). 
The $v_{FWHM}$ (corrected for the spectral resolution) of each fit is listed in each panel and is also summarized in Table~\ref{tab:FWHM_SNhunt120}. 

Comparison of the various line profile fits in the first spectrum reveals a non-negligible difference among the two prevalent Balmer lines.  
In particular, the $H\beta$ feature exhibits a $v_{FWHM}$ that is a factor of $\approx$2 higher than that measured for $H\alpha$, while the Paschen lines exhibit velocities more consistent with $H\beta$. To explain this discrepancy with $H\alpha$, we  performed pseudo-equvialent-width (pEW) measurements in order to estimate the $H\alpha$ to $H\beta$ pEW ratio.
With pEW values of $-53.02$~\AA\ and $-18.75$~\AA\ for $H\alpha$ and $H\beta$, respectively, we obtain a $H\alpha$ to $H\beta$ pEW ratio of 2.83. 
This value is fully consistent with the ratio 2.86  for case~A recombination \citep{osterbrock89}. The $P\beta/H\beta$ and
$P\gamma/H\beta$ ratios are high, indicating that the density is above
the critical density for these high-excitation lines.

The observed velocity difference is likely due to the suppression of
the red line wing by occultation effects commonly observed in
stellar winds \citep[see, e.g.,][and references
therein]{ignace2016}.
The line profiles are quite symmetric, which likely  indicates
optically thick lines in a homologous velocity field \citep[][see their
Fig.~8]{ignace2016}. 
However, it is also possible that we see
rotational broadening from material that was 
in Keplerian orbits with $v = \sqrt{\frac{G M}{r}}$, likely related to an accretion disk.
If this material were ejected by a shock, it is possible that the conservation of angular momentum would lead to conservation of angular velocity. We consider this idea plausible, but 
suspect that the former of optically thick lines in a homologous flow is more probable.

We conclude here with a few words on the presence of helium. The visual wavelength spectra of SNhunt120 shown in Fig.~\ref{Fig:SNhunt120-spectra} lack any discernible features associated with \ion{He}{i} lines. However, as pointed out above, we do link a feature in the $+$41~d NIR spectrum of SNhunt120 to \ion{He}{i} $\lambda$10830 (see Fig.~\ref{fig:lineprofilesLSQ12brd}). Evidence for \ion{He}{i} lines was  previously documented in a series of high-dispersion visual wavelength spectra of the ILRT NGC~300-2008-OT \citep{2009ApJ...699.1850B}, therefore  there is precedent for this, and \ion{He}{i} $\lambda$10830 was also documented in NIR spectra of NGC~300-2008-OT \citep{valerin19}. 
The presence of helium might indicate that a helium-rich envelope is formed through a  second dredge-up phase (see Sect.~\ref{sec:binary?}). However, at $+$40~d the $T_{BB}$ is  6,000~K (see Fig.~\ref{Fig:BB1}, right panel), which is too cool to excite \ion{He}{i} lines, and therefore any signatures of helium are likely due to circumstellar interaction (CSI) and not necessarily due to a high helium abundance.

\subsection{Progenitor constraints}
\label{SNhunt120progenitorconstraints}

Because of the distance of SNhunt120, we 
 cannot reach the expected magnitude level of the various progenitor types, as was done, for example, with SN~2008S \citep[e.g.,][]{2009MNRAS.398.1041B}. In particular, the low resolution of Spitzer (+ IRAC) precludes any attempt of detecting the progenitor in the MIR. We searched public archives for images obtained with ground-based telescopes, but were unable to locate any suitable NIR images; we note that UKIRT (+ WFCAM) images of NGC~5775 are too shallow to be useful.

In order to study the physics of the progenitor,
 we examined a set of HST (+ ACS) images taken on 21 August 2005. NGC~5775 was observed with the $F625W$ filter (approximately equivalent to Sloan $r$ band) for 4$\times$573~s and with the $F658N$ filter (6848~s total split across 12 exposures).
To  determine the location of SNhunt120 in these data, a series of 23 Swope $i$-band images were used to identified 17 sources om common with the HST (+ ACS) $F625W$ filter image. 

We take the average of the transformed positions of SNhunt120 as determined from each of the Swope images as our best estimate of the progenitor location in the prediscovery image, which is plotted in Fig.~\ref{Fig:SNhunt120progenitor}. 
This exercise provides an x,y pixel coordinate for SNhunt120 of $3605.77\pm2.35$ and $1018.15\pm2.29$ (indicated in Fig~\ref{Fig:SNhunt120progenitor} with a dashed circle).
 Here the uncertainties reflect the standard deviation among the transformed positions. 

Using \texttt{DOLPHOT}, a dedicated package for HST photometry\footnote{\url{http://americano.dolphinsim.com/dolphot/}}, we performed PSF photometry on the F625W images. \texttt{DOLPHOT} detected over 50 sources above a 3-sigma significance and a single source that is somewhat consistent with the transformed location of SNhunt120. 
%\LEt{just to be clear: I would remove this entire next sentence because it repeats detail form the figure caption}
The position of sources detected with significance are indicated in Fig.~\ref{Fig:SNhunt120progenitor}, %with red squares, 
while the source (detected with a 17-sigma significance) nearest the expected location of SNhunt120  is also denoted with an A. 
Source A is offset by 3 pixels, or 0\farcs15 from the center of the dashed circle and has 
an apparent F625W magnitude of $m_{F625W} = 25.89\pm0.06$ in the HST flight system (VEGAMAGs). 
In addition, a lower significance (4.1-sigma) source is also detected 3.3 pixels from the center of the circle and is denoted source B, with a brightness of $m_{F625W} = 27.56\pm0.28$ mag.

Finally, we considered the possibility that the progenitor of SNhunt120 was {\it \textup{not}} detected in the ACS image. In order to determine the limiting magnitude, we selected a 4\arcsec\ region surrounding the transient where the background and source density was comparable to that at the progenitor position. 
A comparison of the apparent F625W magnitudes of the extracted sources versus their signal-to-noise ratios reveals a clear peak in the density of sources at $m_{F625W}=27$ mag. This value was adopted as a conservative limiting magnitude. 
This limiting magnitude is shallower than the detection of source A, and it implicitly accounts for realistic possibilities such as two closely blended sources, or local areas of high background.

In the absence of a high-resolution image of the host NGC~5775, we cannot draw strong conclusions from this analysis. If the  source denoted A is the progenitor, then it has an absolute magnitude of $-$4.13$\pm$0.04 mag for our adopted distance and extinction. Otherwise, we set an upper limit to the progenitor absolute magnitude of $F625W > -4.7\pm0.3$ mag (where the uncertainty is due to distance and extinction). Additional circumstellar extinction would serve to mask the progenitor in the optical bands, and without NIR or MIR data, we are therefore unable to provide firm conclusions.

\section{Discussion}

\subsection{Comparison of SNhunt120 to other intermediate-luminosity red transients}
\label{sec:isSNhunt120aILRT}

We plot in Fig.~\ref{fig:SNhunt120_spectracomparison} the  $+0.9$~d and $+$35.7~d  visual wavelength spectra of SNhunt120 compared to a $+$10.5~d spectrum of AT~2017be, a $+16.9$~d spectrum of PTF10fqs, a previously unpublished $+36$~d spectrum of NGC~300-2008-OT1, and a $+$39.2~d spectrum of SN~2008S.  
All of these spectra were corrected for the  Milky Way (MW) and host reddening. 
The spectra of these objects are broadly similar and display the same spectral features (see Sect.~\ref{SNhunt120spectroscopy}) at similar velocities.
Inspection of Fig.~\ref{fig:SNhunt120_spectracomparison} reveals conspicuous H$\beta$ and H$\alpha$ Balmer emission features, as well as features attributed to forbidden [\ion{Ca}{ii}] $\lambda\lambda$7291, 7324 doublet, the \ion{Ca}{ii}  NIR triplet, and \ion{Ca}{ii} H\&K. 
Some differences are apparent, however:  PTF10fqs and NGC~300-OT-1 both exhibit redder continua than the other objects, and they also exhibit weaker features at the blue end of their spectra. 

In Fig.~\ref{Fig:abmagSNhunt120} we compare the absolute magnitude light curves of SNhunt120 with those of M85~2006-OT1, SN~2008S, NGC~300-2008-OT1, PTF10fqs, AT~2017be, and AT~2019abn.
The plotted photometry of these objects has been corrected for reddening and placed on the absolute flux scale adopting the values reported in Table.~\ref{tab:comp_obj_snhunt120} for total color excess and distance modulus.
The majority of these events exhibit a rise to a maximum value typically within their first two weeks of evolution, and overall, they also exhibit similar light-curve decline rates post maximum. 
Interestingly, SNhunt120 is as bright as SN~2008S in all the optical bands. Quantitatively, from comparison of SNhunt120 to other ILRTs, we find that it is $\sim$~2~mag brighter than the faintest objects M85~2006-OT1,  PTF10fqs and  AT~2017be, and about a magnitude fainter than AT~2019abn \citep{jencson2019,at2019abn}, the brightest ILRT  observed to date. 

A common characteristic  of well-observed ILRTs,  for example,  SN~2008S \citep{2009MNRAS.398.1041B}, NGC~300-2008-OT \citep{2009ApJ...705.1425P,2011ApJ...741...37K}, and AT~2017be \citep{at2017be}, and indeed the reason that they are so named, is their emission at red wavelengths.
According to \citet{2011ApJ...741...37K}, the progenitor stars that explode and give rise to these objects  might be 
cool red supergiants of about  9~$M_{\odot}$ , which means that they are some of the most massive AGB stars, and they are therefore referred to as super-AGB stars.
Such stars  are enshrouded in thick dusty stellar winds capable of absorbing even the X-rays produced from the SN shock breakout whose energy is then reemitted by the dust. 
These winds are so dense that dust can 
 reform rapidly even when it is destroyed by the SN radiation field \citep{2011ApJ...741...37K}. 
When they considered broadband observations of SN~2008S extending through MIR wavelengths, \citet[][see their Figs. 6 and 7]{2009MNRAS.398.1041B} documented a double BB SED as early as  $+$17.3~d after explosion. The first hot BB component reproduces emission over the optical wavelengths and is characterized by a BB temperature of $T^{hot}_{BB} \sim 8100$~K, while the second component reproduces the far-NIR through MIR emission and is characterized by  $T^{warm}_{BB} \sim 600$ K. 

Similarly, the Spitzer MIR photometry of SNhunt120 plotted in Fig.~\ref{fig:spitzer} also suggests a second BB  component  at both  early  and late epochs, characterized by $T^{warm}_{BB}$ values of $\approx 800$~K and $\approx 600$~K, respectively. 
We converted the Spitzer  4.5~$\mu$m  flux-point values  into  luminosity using our adopted distance to the host galaxy NGC~5775 (see Sect.~\ref{SNhunt120details}).
Using  the 4.5~$\mu$m flux value at early times computed by interpolating from the first two Spitzer observations to $+$36.8~d, we obtain a $L_{4.5 \mu m}$ value of $1.2\times10^6$ $L_{\odot}$, and $4.2\times10^5$ $L_{\odot}$ on $+$1155~d.
For comparison, at early times, SNhunt120 and SN~2008S exhibit similar MIR luminosity, while at late phases, SNhunt120 appears to be about three times brighter. \citep{2016MNRAS.460.1645A}.

Finally, we note that the BB radius evolution of SNhunt120 is rather constant, similar to that of other ILRTs (see Fig.4 in \citealp{at2018hso}), and is not consistent with the evolution that is typically exhibited by LRNe. In LRNe, the BB radius tends to increase significantly over time (see Paper 2).

\subsection{Context with electron-capture supernovae }
\subsubsection{Comparison of UVOIR light curves of ILRTs to EC SN models}
\label{sec:SNhunt120-ECSN}
Massive stars with zero-age main-sequence (ZAMS) masses $\gtrsim$ 10 $M_{\odot}$ produce degenerate Fe-rich cores that drive the collapse of their inner cores. Depending on the ZAMS mass, such stars  follow one of two paths to their demise. On one path,   a neutron star is formed during core collapse, and subsequently, a  supernova is produced. Alternatively, certain ZAMS mass ranges may follow a second path where their gravitational potential is so great that it is able to overcome the nuclear forces within the proto-NS core and it directly collapses into a black hole. On this second path, the entirety of the star falls back onto the BH and no supernova is produced.  
On the other hand, less massive stars in the range between $\sim$ 8-10 $M_{\odot}$ are expected to form degenerate oxygen-neon-magnesium (ONeMg) cores as they evolve through the super asymptotic giant branch (super-AGB) phase \citep[see][for a recent review]{2017PASA...34...56D}. 
If such ONeMg-core stars approach the Chandrasekhar mass, they may experience  electron-capture reactions that  induce core collapse and the  production of  low-luminosity electron-capture (EC) SNe 
\citep{1980PASJ...32..303M,1984ApJ...277..791N,1984A&A...133..175H,1987ApJ...318..307M,1987ApJ...322..206N} and  neutron star remnants.
An alternative  hypothesis is that instead of undergoing electron-capture-induced collapse, such ONeMg-core stars could experience a partial thermonuclear explosion, leading to the formation of  a ONeFe white dwarf \citep{jrp+16,jrf+19,kjs+19,tj19}. 

A number of studies have suggested a link between ILRTs and EC SNe  
\citep{2009MNRAS.398.1041B,thompson09,2009ApJ...705.1425P,pumo2009,2010Natur.465..326K}. We  examine the UVOIR light curve of SNhunt120 below in relation to the comparison sample of ILRTs and to  two EC SN modeled light curves that exclude and include CSI. This is followed by Sect.~\ref{ECSNmasses} where we present an effort to measure the expected ZAMS mass range of stars that could be the progenitors of EC SNe, followed by Sect.~\ref{sec:binary?} with a short discussion of the possibility that SNhunt120 had a binary origin.

In Fig.~\ref{Fig:bolocompSNhunt120} the UVOIR  light curve of SNhunt120 is compared with those of our ILRT comparison sample and  to synthetic  light curves  calculated by \citet{2014A&A...569A..57M}. 
 These synthetic light curves  combine the EC SN models  presented by  \citet{2013ApJ...771L..12T} with radiative transfer calculations  accounting for  emission produced from CSI. 
It is reasonable to expect CSI  to contribute to the overall energy emission budget in at least some  super-AGB stars as they are commonly enshrouded in dense CSM \citep[see][]{2017PASA...34...56D}. We note,  however, that a number of factors influence the post-carbon burning evolution and hence the  circumstellar environment, including the initial mass and the composition, the pre-explosion mass-loss history, and the core-growth efficiency \citep[e.g.,][]{pumo2009,2017PASA...34...56D}. We therefore approach   this particular comparison with caution.

Inspection of Fig.~\ref{Fig:bolocompSNhunt120} reveals that consistent with the absolute-magnitude light-curve comparisons in Fig.~\ref{Fig:abmagSNhunt120}, SNhunt120 lies at the bright end of the peak luminosity distribution.
At peak, SNhunt120 is as bright as SN~2008S, which until the recent discovery of AT~2019abn \citep{jencson2019,at2019abn}
 was the record holder for being the brightest ILRT. Compared to the faintest ILRTs, SNhunt120 and SN~2008S are  a factor of  $\sim$10 more luminous, with the added caveat that there might be even fainter ILRTs that have remained undiscovered due to selection biases. 
 These findings follow what is demonstrated in Fig.~\ref{Fig:Mr_vs_time}, which also provides a perspective on where ILRTs and LRNe classes lie within the parameter space of peak luminosity versus decay time. 

The EC SN synthetic light curve models presented by \citet{2014A&A...569A..57M} that we plot in Fig.~\ref{Fig:bolocompSNhunt120} (with and without CSI) reach peak luminosities of log$_{10}(L_{\rm peak}) \approx 42.1$ erg~s$^{-1}$. This makes them approximately a factor of  $\sim$15 brighter than SNhunt120 and SN~2008S. Although the explosion energy is predicted to be low (around $10^{50}~\mathrm{erg}$) in EC SNe \citep{2006A&A...450..345K}, the early bright luminosity in the EC SN light-curve model is caused by the extended envelope and low envelope mass found in super-AGB SN progenitors \citep{2013ApJ...771L..12T}. If ILRTs are actually EC SNe, the explosion energy of EC SNe needs to be $\sim 10^{48}~\mathrm{erg}$ or lower to have the lower observed peak luminosity \citep{2013ApJ...771L..12T}. 
This in turn suggests that current EC SN explosion simulations overpredict the explosion energy. However, in principle, the envelope mass might  be higher by as much as a factor of 2--3 than what was assumed in the \citet{2013ApJ...771L..12T} and \citet{2014A&A...569A..57M} models. If this is indeed the  case, we might  expect the peak luminosity to be affected, and therefore this should also be considered in future modeling efforts. 

\subsubsection{Mass range of  EC SN progenitors}
\label{ECSNmasses}

The expected range of  ZAMS masses of isolated stars producing EC SNe is  expected to be rather confined, and depending on metallicity,  to be about $0.1-0.2\;M_\odot$ \citep[][and references therein]{2017PASA...34...56D}. However, within a ZAMS interval of roughly $8-11\;M_\odot$ \citep{jhn+13,jhn14,dgs+15,wh15}, the exact location of this confined parameter space for EC SN progenitors could be shifted by $1.5-2.5\;M_\odot$, depending on the metallicity and core-convective overshooting \citep[e.g.,][]{siess2006,sie07,2008ApJ...675..614P}.

Assuming a Salpeter initial mass function (IMF) with isolated single stars as progenitors, 
we can compute the expected range of ZAMS masses for the progenitors of EC SNe given the known observed fraction of 2008S-like events relative to the rate of core-collapse SNe (10\%--20\%, from \citealp{thompson09}). This value is probably too high because that ILRTs and LRNe were not considered separate objects at the time of \citet{thompson09}. 
If the lowest ZAMS mass to produce an EC SN is 8.5~$M_{\odot}$, then the highest ZAMS mass will be between 9.1 and 9.7 $M_{\odot}$, assuming a Salpeter IMF with isolated single stars as progenitors. 

On the other hand, assuming that the lowest ZAMS mass of a progenitor star to produce an EC SN is 8.0~$M_{\odot}$, then the highest ZAMS mass will range between $8.6-9.1$ $M_{\odot}$. Finally, when we assume a lower ratio of EC SNe over CC SNe, for example, 5\%, as discussed in \citet{2009MNRAS.398.1041B}, the range of progenitor masses would be 8.0--8.3~$M_{\odot}$ or 8.5--8.8~$M_{\odot}$ depending on the assumption on the lowest ZAMS mass.

In a similar vein, taking the mass range of the EC SN channel from \citet{poe07} and weighting it with a Salpeter IMF, \citet{2017PASA...34...56D} estimated  that the EC SN contribution to the overall core-collapse SN rate is 5\%, 17\%, and 38\% of all type~IIP~SNe for metallicities $Z = 0.02, 0.001,$ and $10^{-5}$, respectively, when the maximum mass for a type~IIP~SN is assumed to be $18\;M_\odot$, based on the analysis of SN observations by \citet{sma15}.
In comparison with our estimate above, we find good agreement with their result for $Z=0.001$.
These results are, however, in contrast to 
findings of \citet{2017PASA...34...56D}.
Specifically, when the results from  \citet{dgs+15} at these same metallicities were used, \cite{2017PASA...34...56D} calculated a 
far lower percentage of as  low as 2--5\% that all type~IIP~SNe progenitors will lead to an EC SNe.
The main reason for the variation in frequency of EC SNe is due to the uncertainty in constraining the mass-loss rate of stars at low metallicity.  Moreover, the EC SN candidates found in the local Universe (including SNhunt120) are all at approximately solar metallicity.

\subsubsection{Binary origin for SNhunt120?}
\label{sec:binary?}
It has previously been demonstrated that the width of the initial mass range for producing EC SNe is smaller in single stars than in binary stars in nondegenerate systems \citep{plp+04,2008ApJ...675..614P}. The reason for this effect is that the presence in a binary can dramatically affect the structure of the core of a massive star at the time of core collapse. Stars above $\sim 11\;M_\odot$ are generally expected to have smaller iron cores if they lose their envelopes in a close binary. Stars in the range of $8-11\;M_\odot$ may explode as EC SNe if they are located in a {\em \textup{close}} binary, while stars in {\rm wide} binaries or single stars will experience a 
second dredge-up phase and are more likely to end their evolution as ONeMg white dwarfs.
 \citet{plp+04} argued that the minimum initial mass of a massive single star that becomes a neutron star may be as high as $10-12\;M_\odot$, while for close binaries, this mass may be as low as $6-8\;M_\odot$. These critical masses depend on the applied input physics. 
This includes the treatment of convection, the amount of convective overshooting, and the metallicity of the star, and will generally be lower for larger amounts of convective overshooting and lower metallicity \citep{plp+04,siess2006}.

Although a large portion of main-sequence massive stars are found in close binaries \citep{sdd+12}, the hydrogen-rich spectra of SNhunt120 point to an isolated, or wide-orbit, progenitor star. 
In contrast, SN~Ibn progenitors have lost their hydrogen (and  also part of their helium) envelopes and
are embedded in dense helium-rich CSM, giving rise to narrow helium spectral features that appear in visual wavelength spectra \citep[e.g.,][]{2017ApJ...836..158H}.

\section{Conclusion}
\label{sec:conclusions}

\begin{figure}
\centering
\includegraphics[width=18cm]{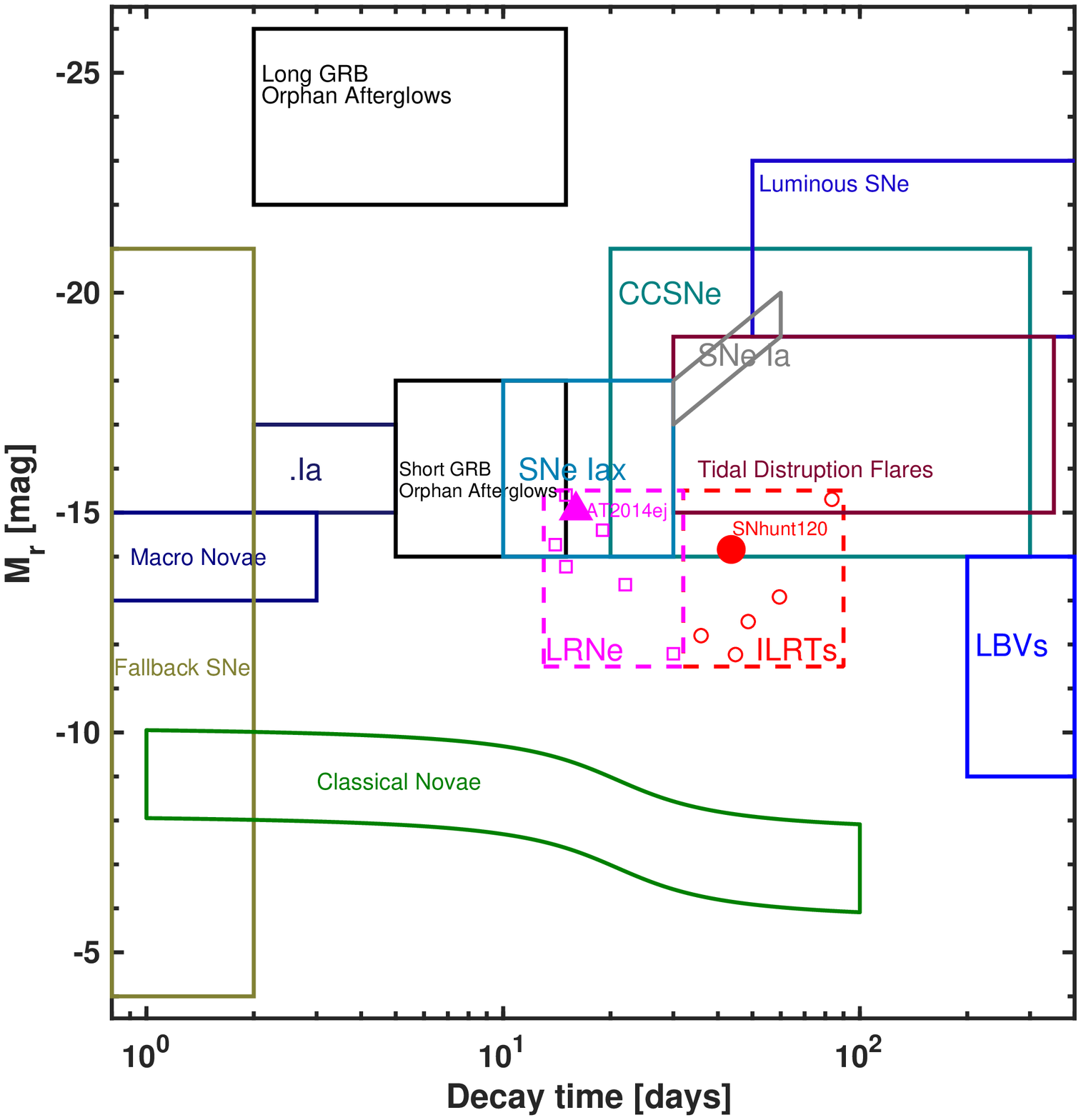}
\caption{Absolute peak $r$-band magnitude vs. decay time (defined as the time taken for the $r$-band light curve to decline one magnitude from peak) for optical transients. The figure is adapted from \citet{kasliwal_phd} and \citet{rau09}. Peak absolute $r$-band  magnitudes and decay times for the ILRTs and LRNe are plotted as red circles and magenta squares, respectively. The LRN AT~2014ej plotted as a magenta triangle is a lower limit (see Paper 2). The LRNe and ILRTs regions are plotted as dashed lines to highlight the point that the full extent of the parameter space occupied by  each subtype is unknown.}
\label{Fig:Mr_vs_time}
\end{figure}

\begin{figure}[!h]
\centering
\includegraphics[width=18cm]{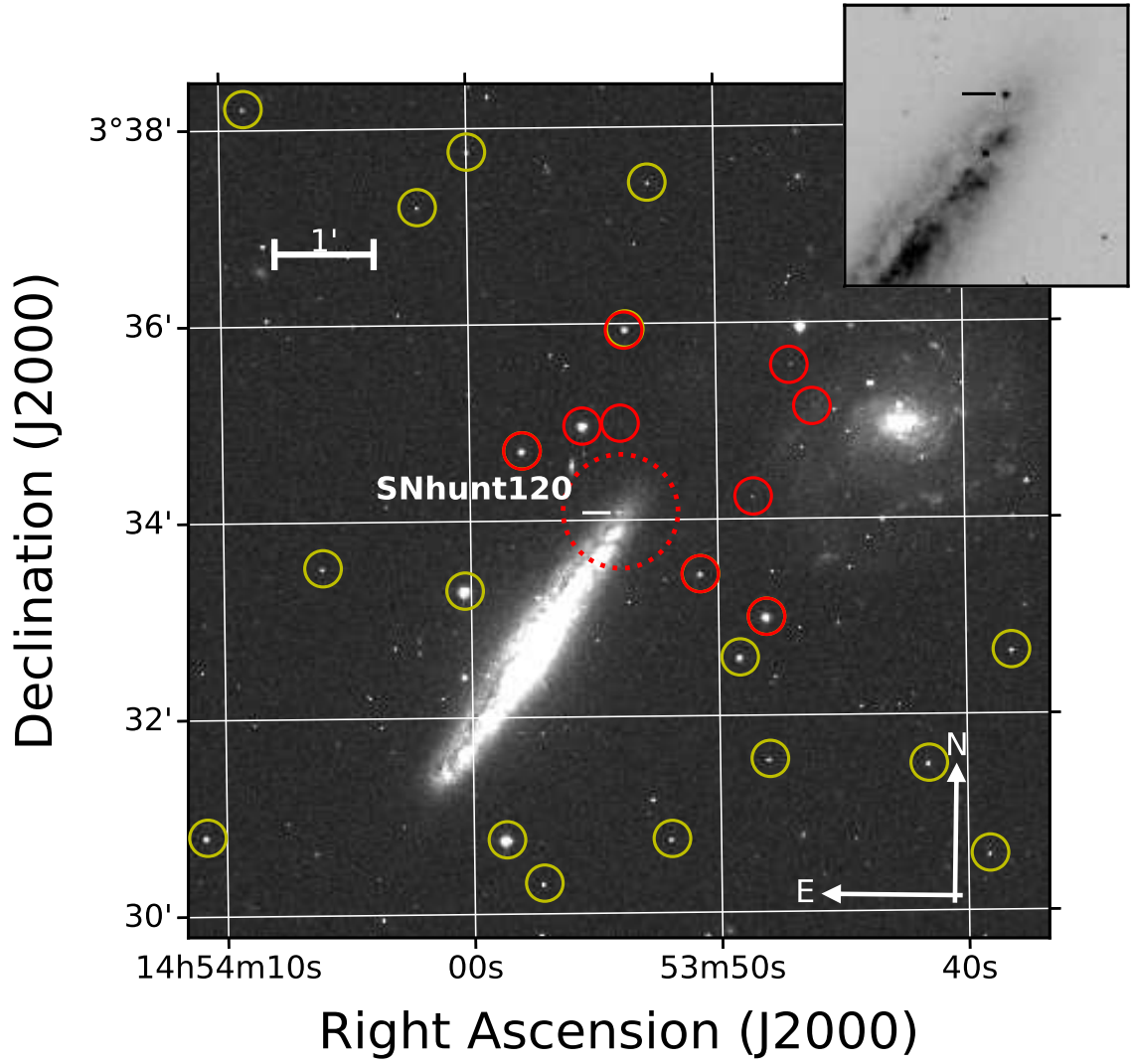}
\caption{ Finding chart of  NGC~5775 hosting SNhunt120  constructed from a single Swope $r$-band image when the transient was near maximum brightness. 
The position of SNhunt120  is indicated with a dotted circle with a zoom into the area shown in the upper right corner inset. Optical and NIR local sequence stars are indicated with yellow and red circles, respectively.}
\label{fig:FC1}
\end{figure}

\begin{figure}
\centering
\includegraphics[width=17cm]{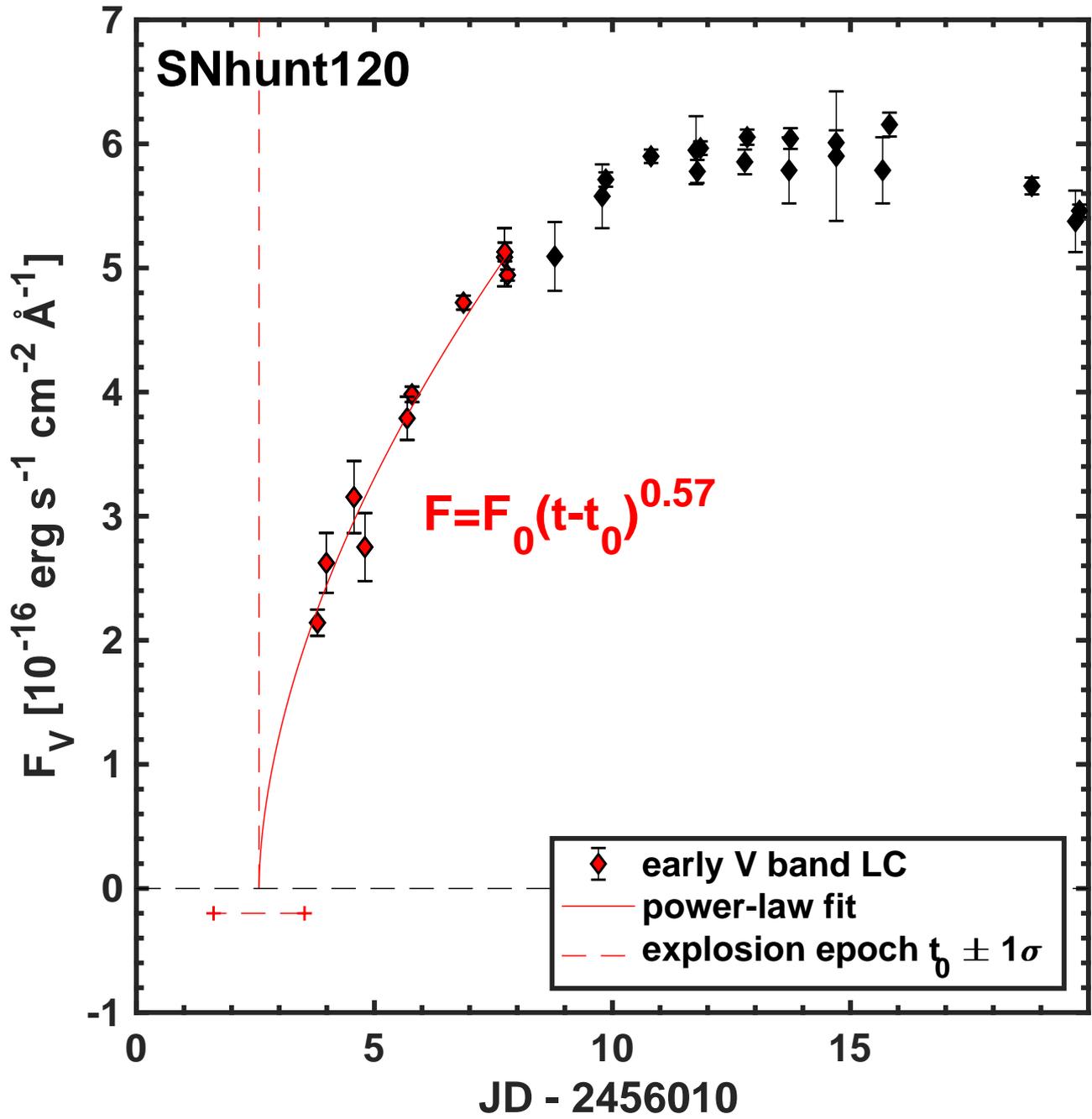}
\caption{Early light curve of SNhunt120 plotted with a  best-fit power-law function (solid line, reduced $\chi^2 = 4.6$) to the first ten epochs (red points) of photometry. 
 The explosion epoch, $t_0$, is  estimated to have occurred on JD+2456012.58$\pm$0.95, which is 1.25 days prior to the first-detection epoch. The dashed red line marks $t_0$ as obtained from the fit, its uncertainty is marked by a red segment just below the zero-flux level.}
\label{LSQearly}
\end{figure}

\begin{figure}
 %  \centering
    \includegraphics[width=9cm]{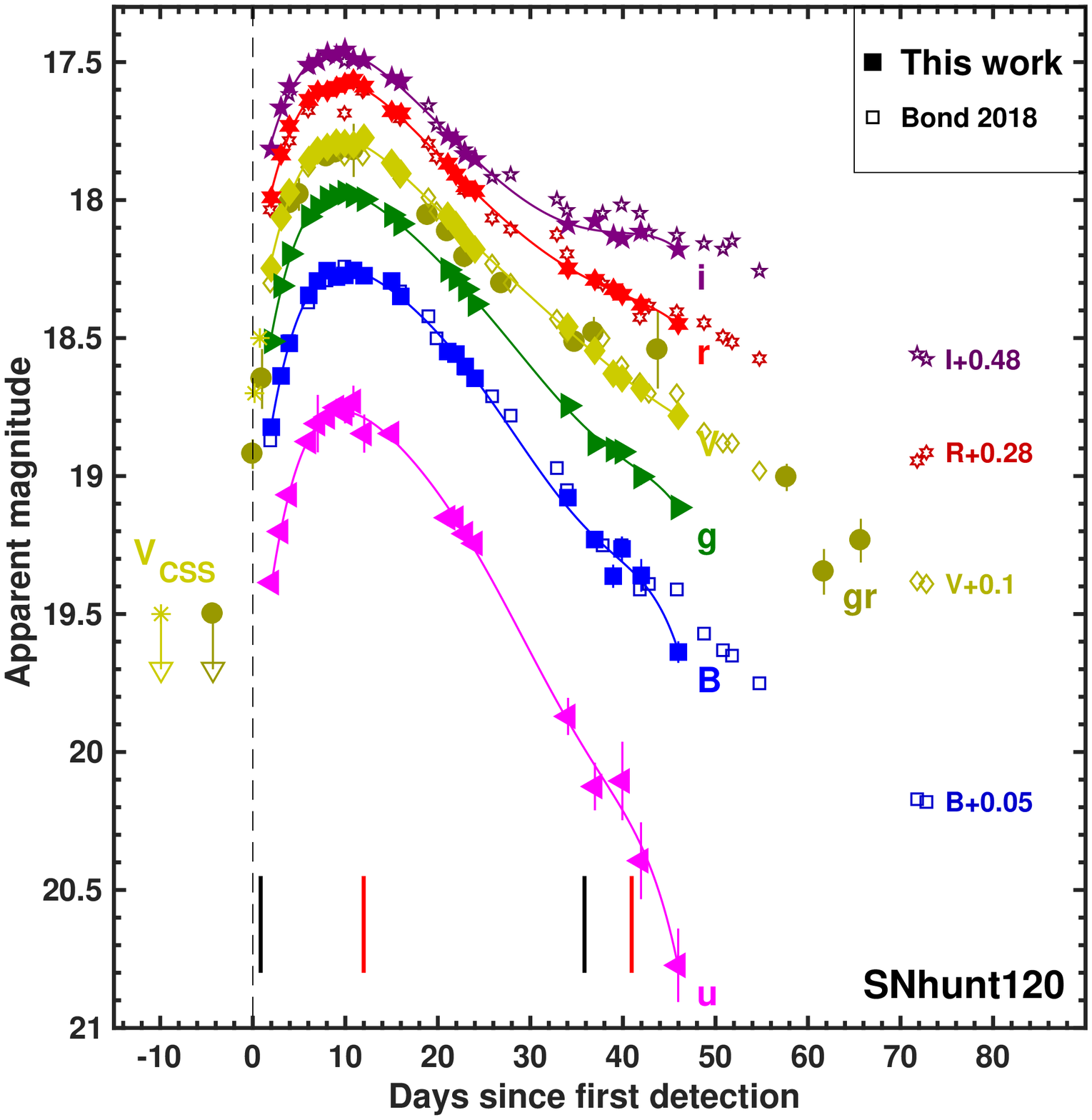}

   \includegraphics[width=9cm]{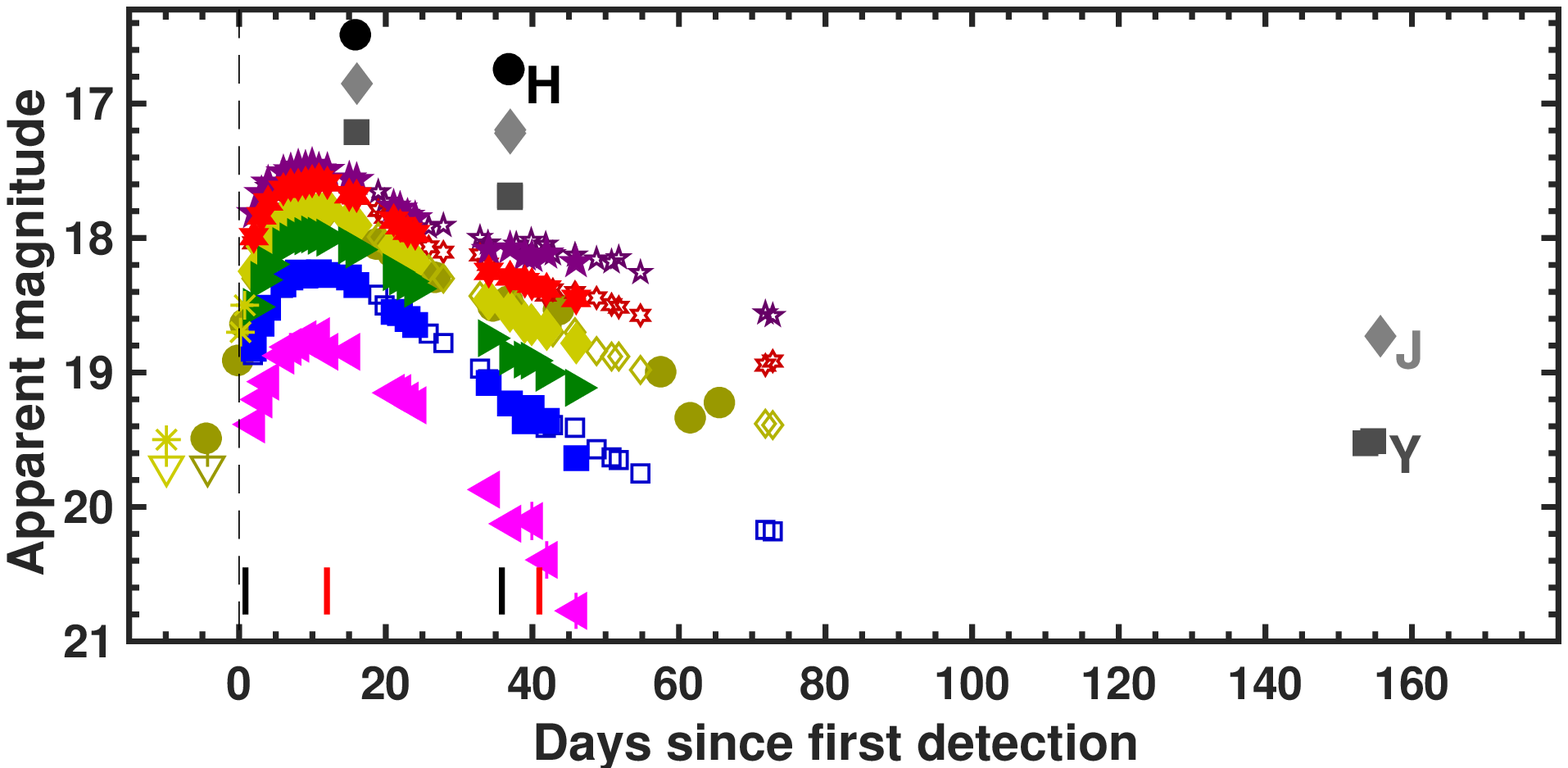}
   \caption{\textit{Top:} optical $uBgVri$-band photometry of SNhunt120 obtained by the CSP-II and  $gr$-band photometry from LSQ. The light curves cover the rise to maximum and subsequent decline over the period of a month and a half. The Catalina Sky Survey (CSS; \citealt{2012ATel.4004....1H}) reported $V$-band discovery magnitude of 18.7 obtained on March 27.49,  a CSS  $V$-band nondetection limit of 19.5 mag obtained from data taken on March 17.39 UT, and an unfiltered photometry point obtained on  March 28.07  of 18.5 mag are also plotted. The $BVRI$  photometry published by  \citet{2018RNAAS...2c.176B} is overplotted with open symbols; each band is shifted by arbitrary constants to match their most similar CSP-II band. The CSP-II light curves are fit with low-order polynomials plotted as solid colored lines.
   Optical spectral phases are marked by black segments, and NIR spectral phases are marked by red segments. \textit{Bottom:} As in the top panel, but also including the NIR ($YJH$) points of SNhunt120.}
   \label{Fig:SNhunt120photometry}
    \end{figure}

We have added  to the growing populations of intermediate-luminosity red transients (ILRTs).  Here in Paper~I we considered SNhunt120, a gap transient discovered within days of its inferred explosion epoch that exhibits many observational properties  consistent with those of other well-studied ILRTs, including an  MIR excess extending from early to late phases. Reaching an $r$-band peak magnitude of $-$15.05~mag and exhibiting a decay time of 44 days, SNhunt120 is one of the brightest and fastest evolving ILRTs observed to date (see Fig.~\ref{Fig:Mr_vs_time}). 
Furthermore, comparison of the UVOIR light curves of SNhunt120 and other  ILRTs to current electron-capture SN simulations reveals that the models overpredict the explosion energy. This  should be considered in future modeling efforts. 
In Paper~2  we present a detailed examination of the luminous red nova designated AT~2014ej.

\begin{acknowledgements}
We thanks the referee for a very constructive report.
A special thanks to David Rabinowitz and Peter Nugent for providing LSQ images of SNhunt120 (LSQ12brd). We also thank Nathan Smith for provided published spectra in electronic format of a handful of  gap transients and Massimo Della Valle for a handful of interesting comments.
The CSP-II has been funded by the USA's NSF under grants AST-0306969, AST-0607438, AST-1008343, AST-1613426, AST-1613455, AST-1613472 and also in part by  a Sapere Aude Level 2 grant funded by the Danish Agency for Science and Technology and Innovation  (PI M.S.).
M.S., F.T. and E.K. are supported by a project grant (8021-00170B) from the Independent Research Fund Denmark (IRFD). Furthermore, M.S. and  S.H. are supported  in part by a generous grant (13261) from VILLUM FONDEN. 
M.F. is supported by a Royal Society - Science Foundation Ireland University Research Fellowship.
T.M.T.\ acknowledges an AIAS--COFUND Senior Fellowship funded by the European Union’s Horizon~2020 Research and Innovation Programme (grant agreement no~754513) and Aarhus University Research Foundation. 
N.B.S. acknowledges support from the NSF through grant AST-1613455, and through the Texas A\&M University Mitchell/Heep/Munnerlyn Chair in Observational Astronomy.
L.G. is funded by the European Union's Horizon 2020 research and innovation programme under the Marie Sk\l{}odowska-Curie grant agreement No. 839090. This work has been partially supported by the Spanish grant PGC2018-095317-B-C21 within the European Funds for Regional Development (FEDER).
Support for J.L.P. is provided in part by FONDECYT through the grant 1191038 and by the Ministry of Economy, Development, and Tourism’s Millennium Science Initiative through grant IC120009, awarded to The Millennium Institute of Astrophysics, MAS.
M.L.P. is partially supported by a ``Linea 2'' project (ID 55722062134) of the Catania University.
This research has made use of the NASA/IPAC Extragalactic Database (NED), which is operated by the Jet Propulsion Laboratory, California Institute of Technology, under contract with the National Aeronautics and Space Administration. We acknowledge the usage of the HyperLeda database.
\end{acknowledgements}

%\clearpage
  \bibliographystyle{aa} % style aa.bst
 % \bibliography{bibliograph.bib} % 

\clearpage

\clearpage

\clearpage

   \clearpage
 
  \clearpage
\begin{figure}
\centering
\includegraphics[width=12cm]{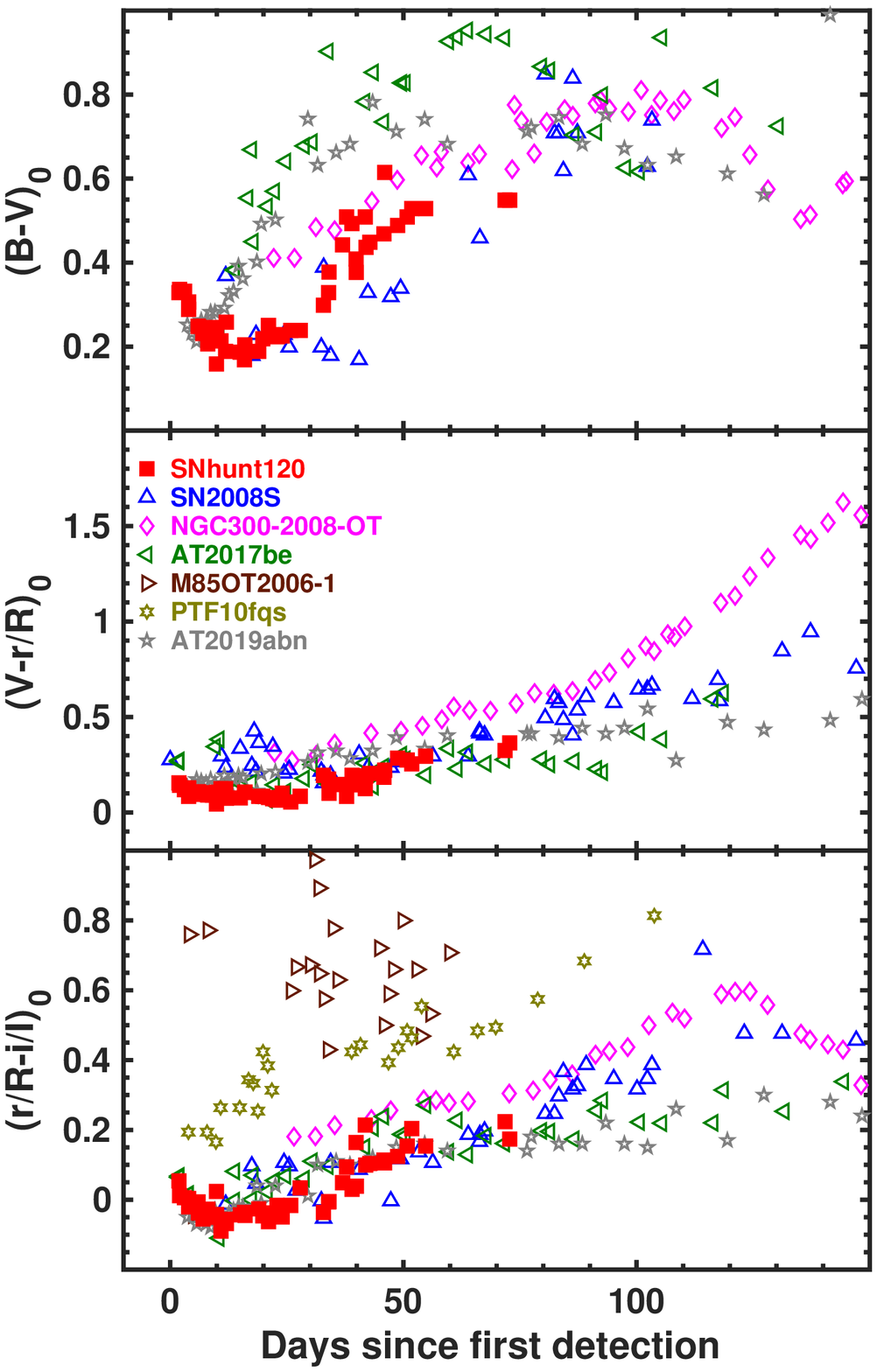}
\caption{Intrinsic broadband optical color evolution of SNhunt120 compared to the  ILRTs M85-OT2006-1 \citep{m85_kulkarni,m85_pastorello}, SN~2008S \citep{2009MNRAS.398.1041B}, NGC~300-2008-OT1 \citep{bond_ngc300ot}, PTF10fqs \citep{kasliwal11}, AT~2017be \citep{at2017be}, and AT~2019abn \citep{at2019abn}. The colors are corrected for reddening adopting the values listed in Sect.~\ref{SNhunt120photometry}. For SNhunt120, we adopt $A^{host}_{V} = 0.64$ mag and a \citet{NED_extinction} reddening law so that its broadband colors match the  colors of the bluest objects in the comparison sample.}. 
\label{Fig:colorSNhunt120}
\end{figure} 

\clearpage
\begin{figure}
\centering
\includegraphics[width=18cm]{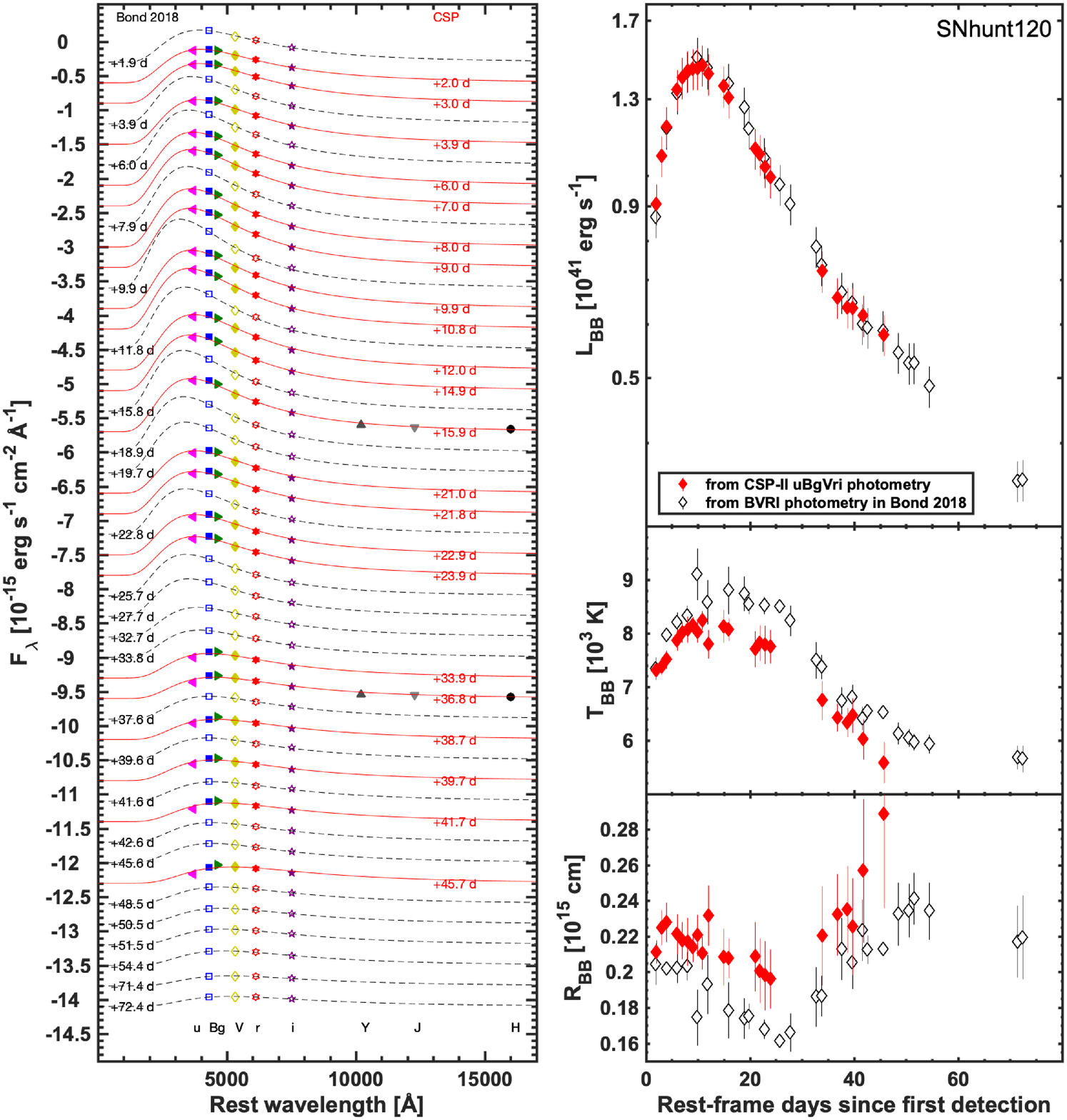}
\caption{\textit{Left:} Spectral energy distributions of SNhunt120 constructed from our optical photometry  and complemented with photometry from \citet{2018RNAAS...2c.176B}. The best-fit  BB is overplotted on each SED function; red lines are the fit to CSP-II data, and black lines are the fit to \citeauthor{2018RNAAS...2c.176B} data.
For clarity, the SEDs are shifted by arbitrary constants, and the phase of each SED relative to the epoch of discovery is indicated. \textit{Right:}  Resulting time evolution of the BB luminosity \textit{(top)}, BB temperature \textit{(middle),} and BB radius \textit{(bottom)} of the underlying emission region corresponding to the BB fits.} 
\label{Fig:BB1}
\end{figure}

\clearpage
\begin{figure}
\centering
\includegraphics[width=12cm]{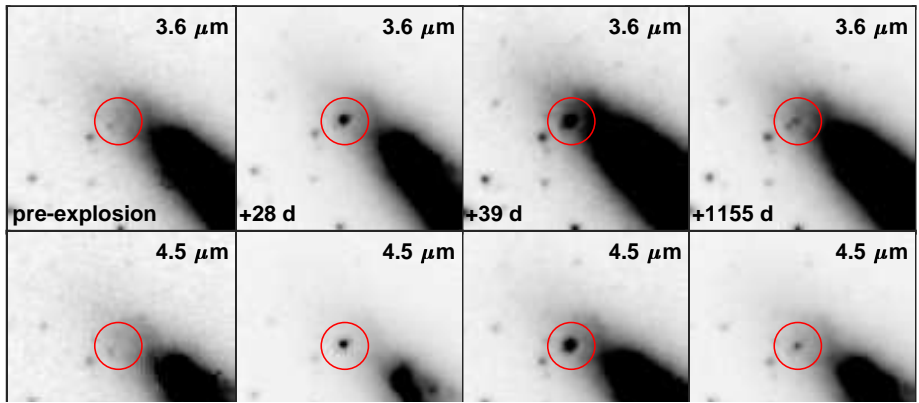}

\includegraphics[width=12cm]{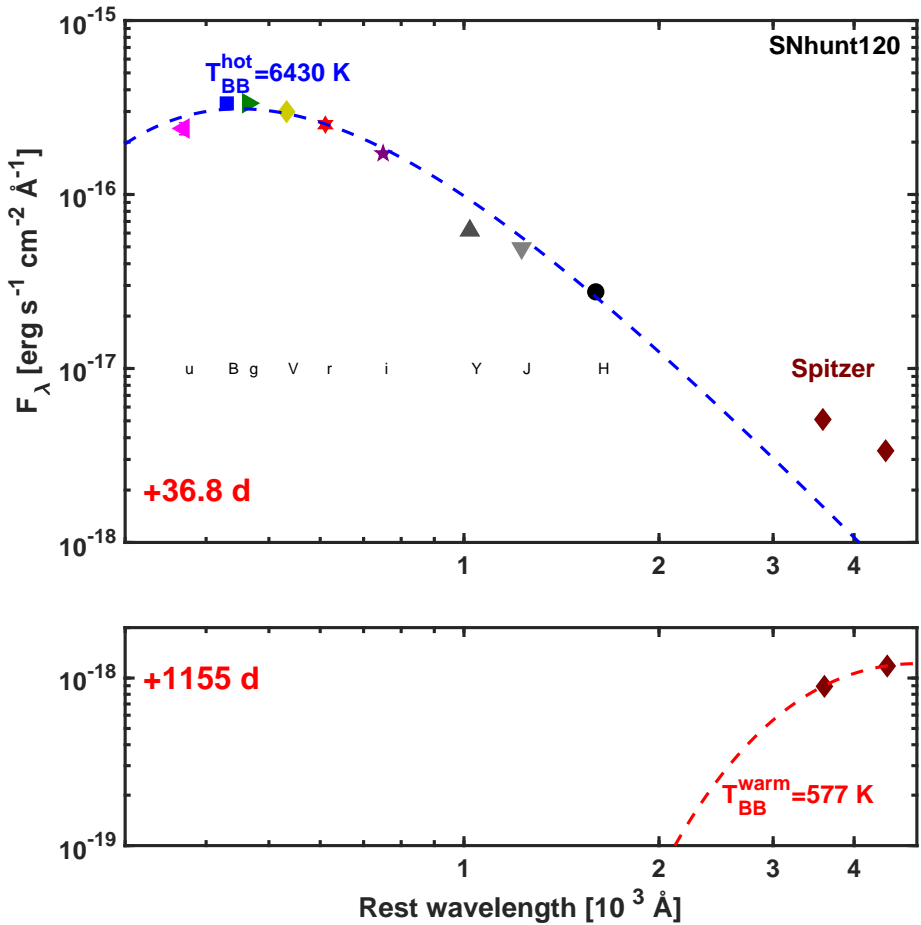}
\caption{ \textit{Top:} \textit{Spitzer} Space Telescope images of SNhunt120 obtained on $+$28~d, $+$39~d, and $+$1155~d at 3.6~$\mu$m and 4.2~$\mu$m, the position of the transient is indicated with a red circle. \textit{Bottom:} Two-panel plot containing SEDs constructed for SNhunt120 at early  and late phases, along with BB fits. The early SED is constructed with CSP-II photometry taken on +36.8~d and Spitzer measurements obtained by interpolating between estimated magnitudes obtained from  $+$28~d and $+$39~d images. A BB function fit (blue dashed line) characterized by a temperature of $T_{BB}^{hot}=6430\pm266$ K is overplotted on the early SED. The BB reproduces the fluxes from $u$ to $H$ band. However, the Spitzer MIR fluxes are clearly above the flux level of the best BB fit, suggesting the presence of an additional cooler component. The SED constructed using the +1155~d Spitzer flux points is fit with a BB function characterized by  $T_{BB}^{warm} \sim 600$~K.} 
\label{fig:spitzer}
\end{figure} 

\clearpage
\begin{figure}
\centering
\includegraphics[width=13cm]{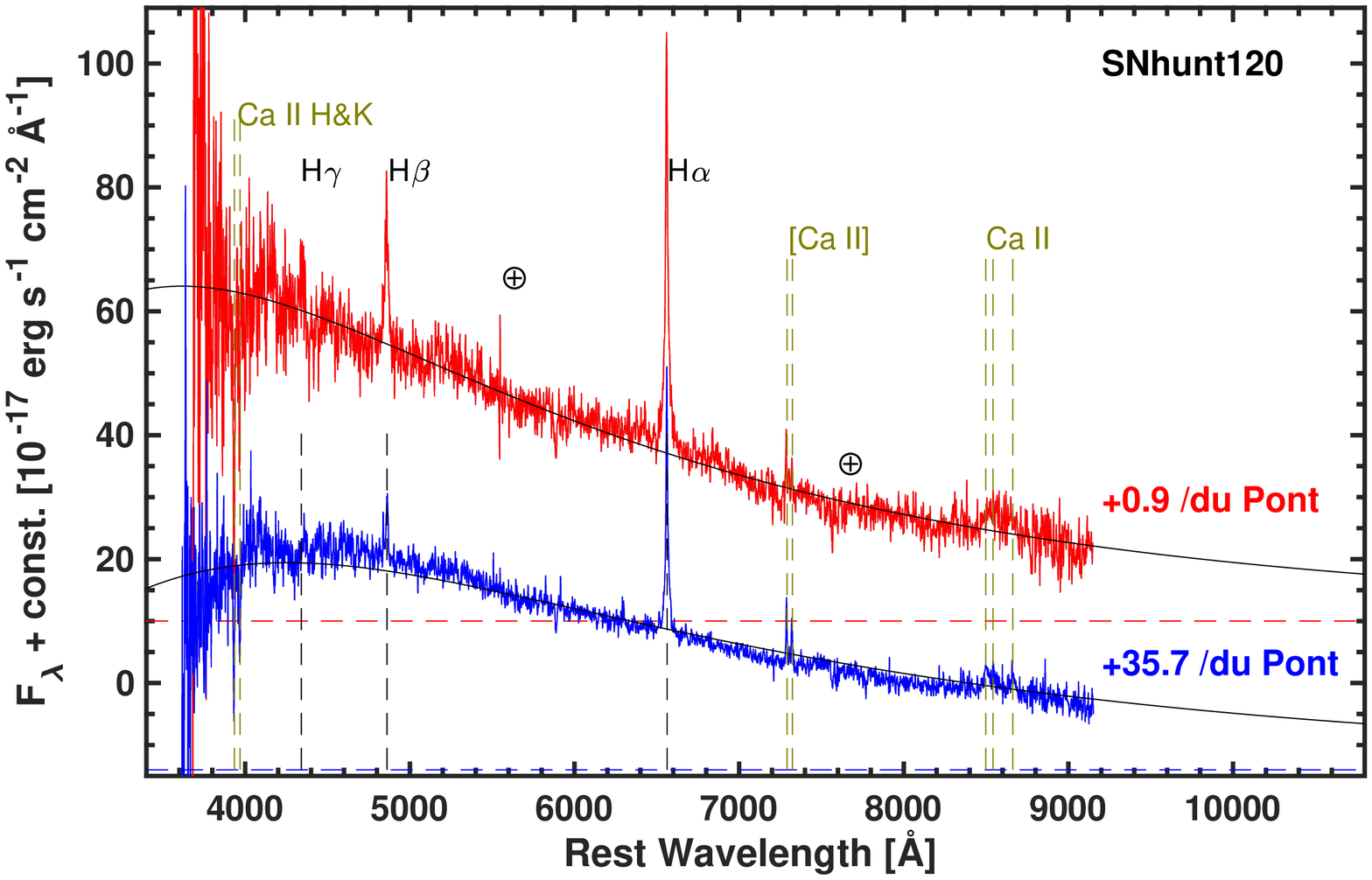}  

%\vspace{1cm}
\includegraphics[width=13cm]{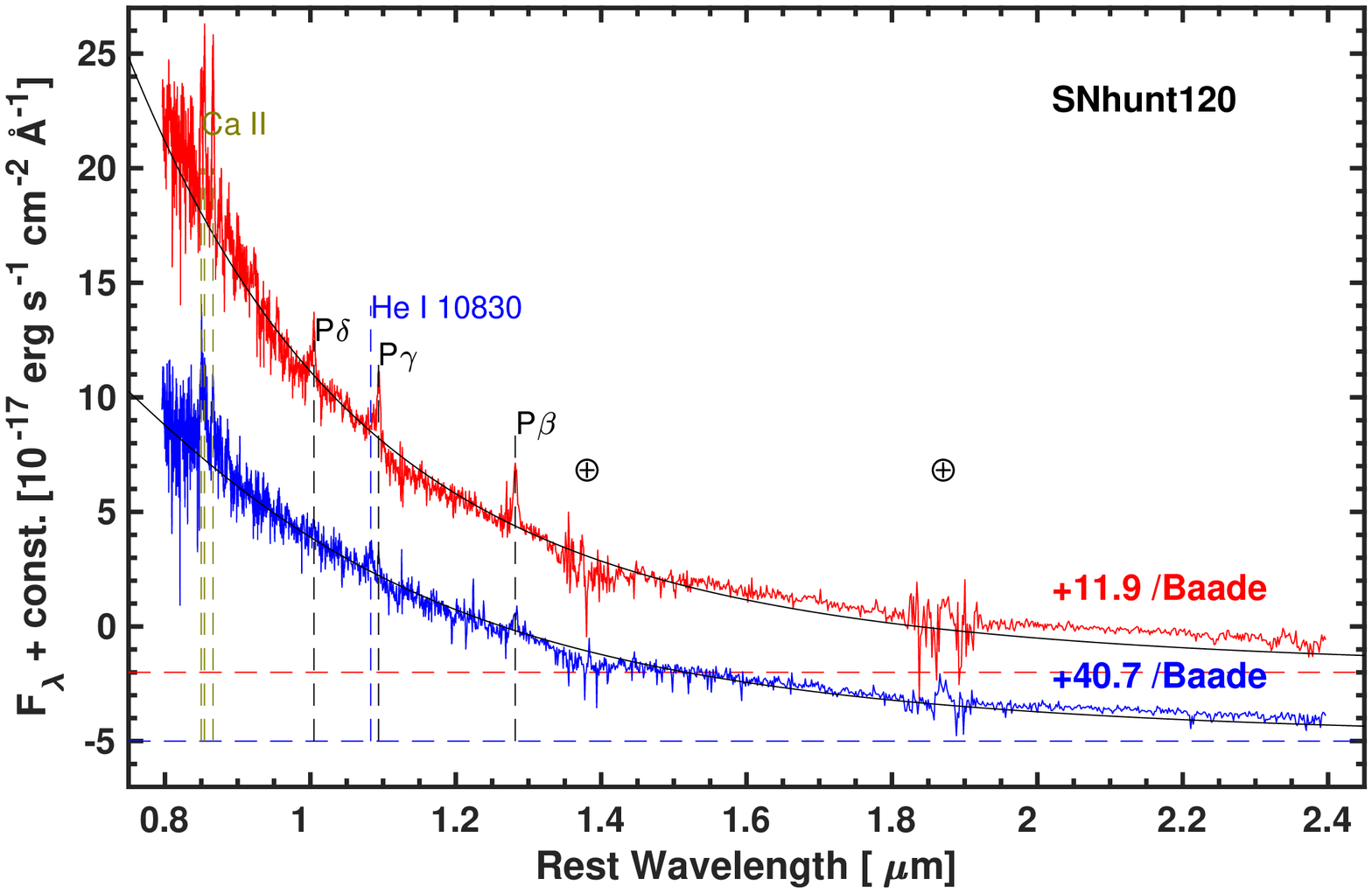}
\caption{Low-resolution visual (top panel) and NIR wavelength (bottom panel) spectra of SNhunt120 in the rest frame. The flux calibration of each spectrum is adjusted to match the corresponding broadband photometry, and the spectra are corrected for both MW and  host-galaxy extinction. The spectra are separated for presentation by the addition of an arbitrary constant. Best-fit BB functions are overplotted as black lines, and the phase of each spectrum relative to the discovery epoch and telescope used to obtain the data is indicated.}
\label{Fig:SNhunt120-spectra}
\end{figure}  

\clearpage
       \begin{figure}
      \centering
\includegraphics[width=18cm]{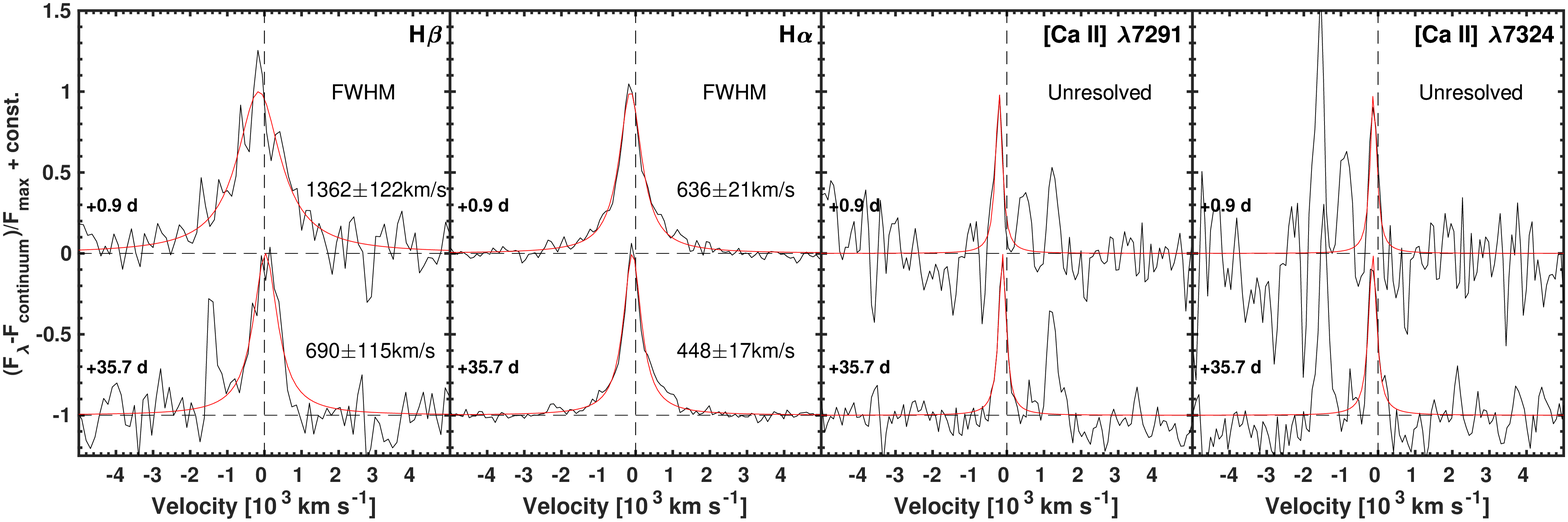}
\includegraphics[width=18cm]{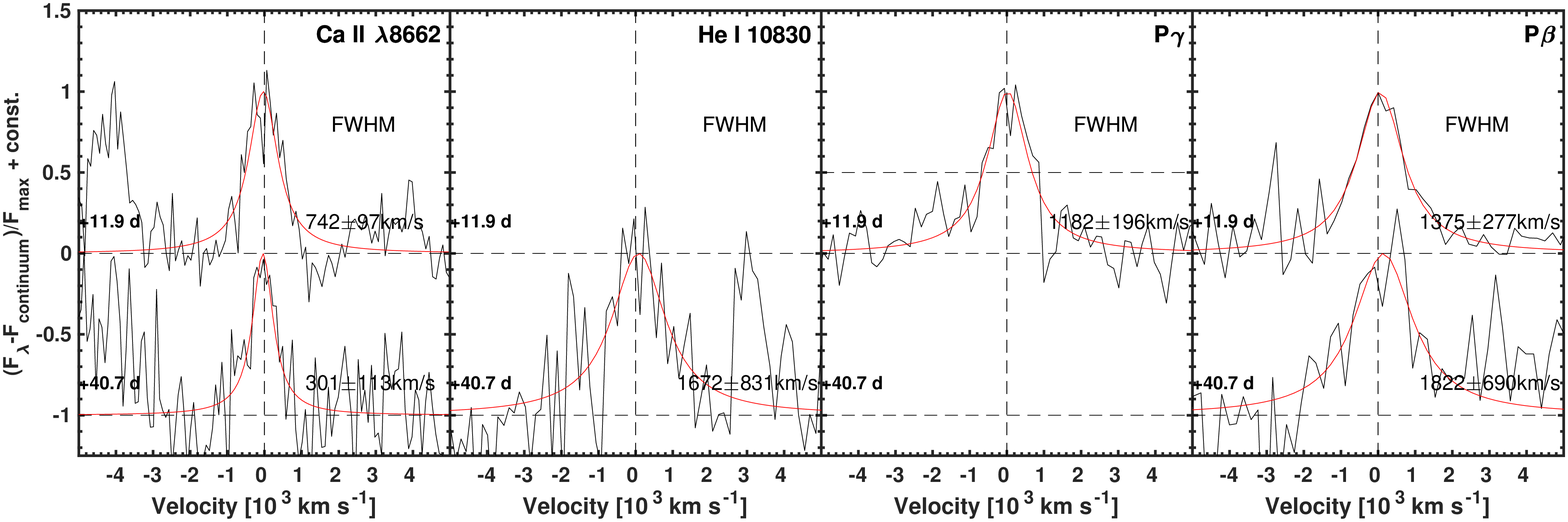}
\caption{Prominent spectral features identified in the spectra of SNhunt120. These include H$\beta$, H$\alpha$, [\ion{Ca}{ii}], the \ion{Ca}{ii} NIR triplet, \ion{He}{i}~$\lambda$10830, P${\gamma}$, and P${\beta}$. The  continuum of each spectrum was subtracted  using a low-order polynomial fit to the observed data, and for presentation purposes, the peak of each line was normalized to unity.
The vertical dotted lines mark the rest velocity position of each ion. Lorentzian line profile fits are overplotted as red lines, and the corresponding line velocities, corrected for the resolution, are provided in the insets and are also listed in Table~\ref{tab:FWHM_SNhunt120}. The forbidden [\ion{Ca}{ii}] lines are unresolved.}
\label{fig:lineprofilesLSQ12brd}
\end{figure}

\clearpage
    \begin{figure}
   \includegraphics[width=18cm]{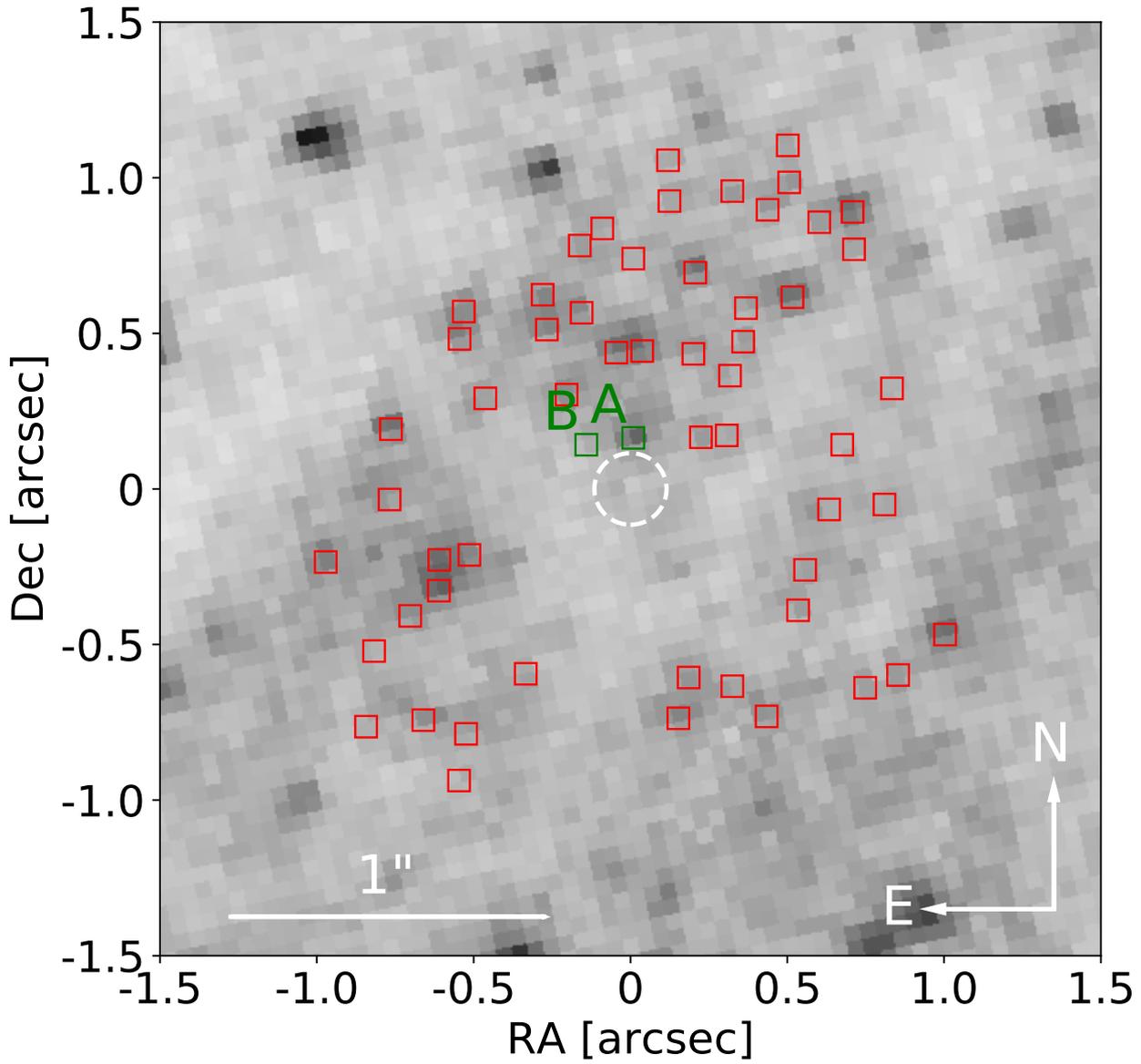}
   \caption{Pre-explosion HST (+ ACS) F625W-band image of NGC~5775. The location of SNhunt120 in the image is highlighted with  a dashed circle with a radius that corresponds to the 1-sigma positional uncertainty. No candidate sources are detected within the dashed circle down to a conservative limiting magnitude of  $m_{F625W}=27$ mag. The green squares indicated with A and B are two nearby sources discussed in the text. The red squares are all sources detected by \texttt{DOLPHOT} above a 3-sigma significance.}
  \label{Fig:SNhunt120progenitor}
    \end{figure}

\clearpage
   \begin{figure}
   \centering
   \includegraphics[width=18cm]{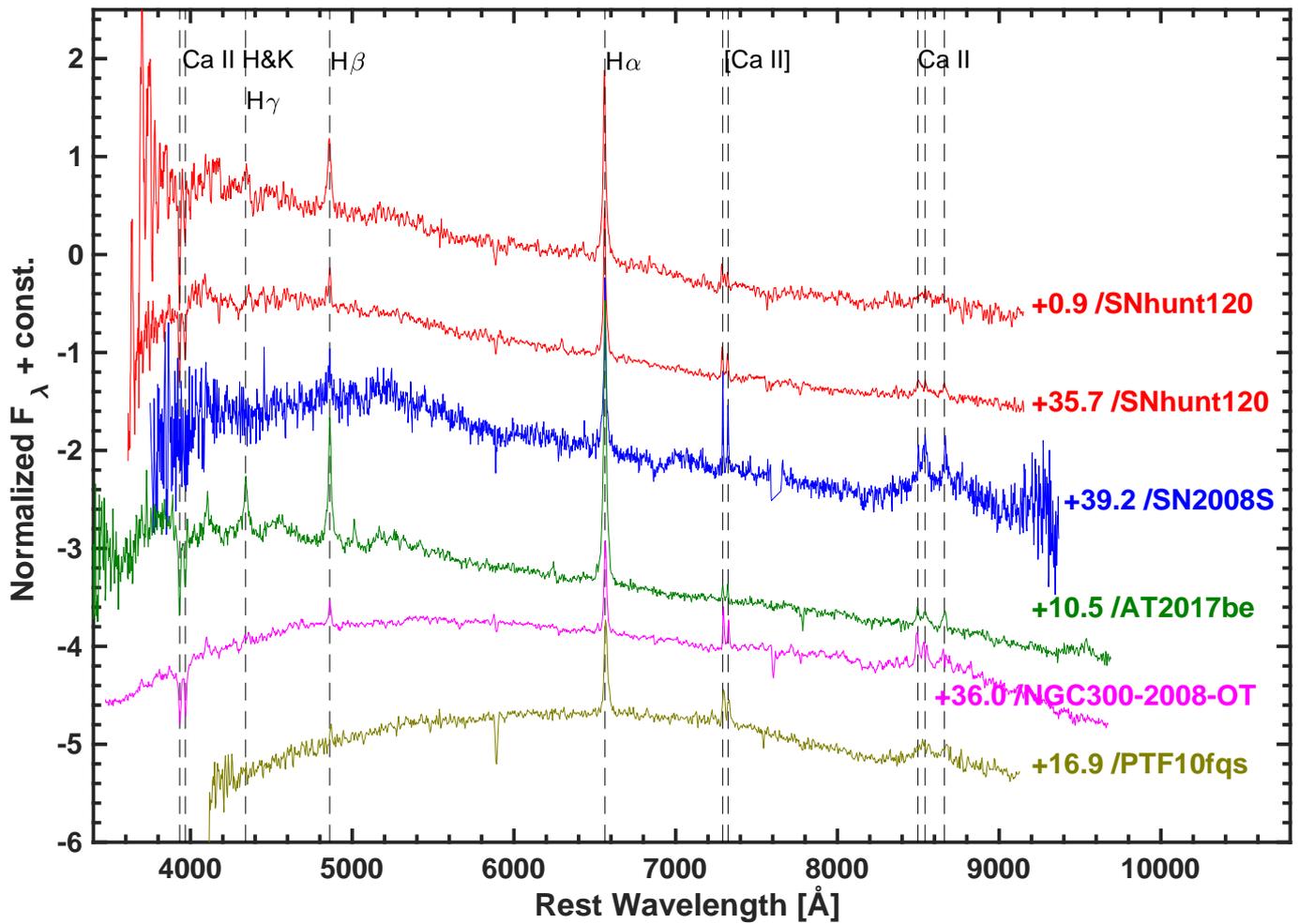}
   \caption{Comparison of our visual wavelength spectra of SNhunt120  with similar phase (since first detection) spectra of  other ILRTs. This includes AT~2017be on $+$10.5~d, PTF10fqs on $+16.9$~d, NGC~300-2008-OT1 (previously unpublished and obtained by CSP-I) on $+$36.0~d, and SN~2008S on $+39.2$~d. Each of the spectra was corrected for reddening.}
      \label{fig:SNhunt120_spectracomparison}
   \end{figure}
   
 \clearpage
\begin{figure}
\centering
\includegraphics[width=18cm]{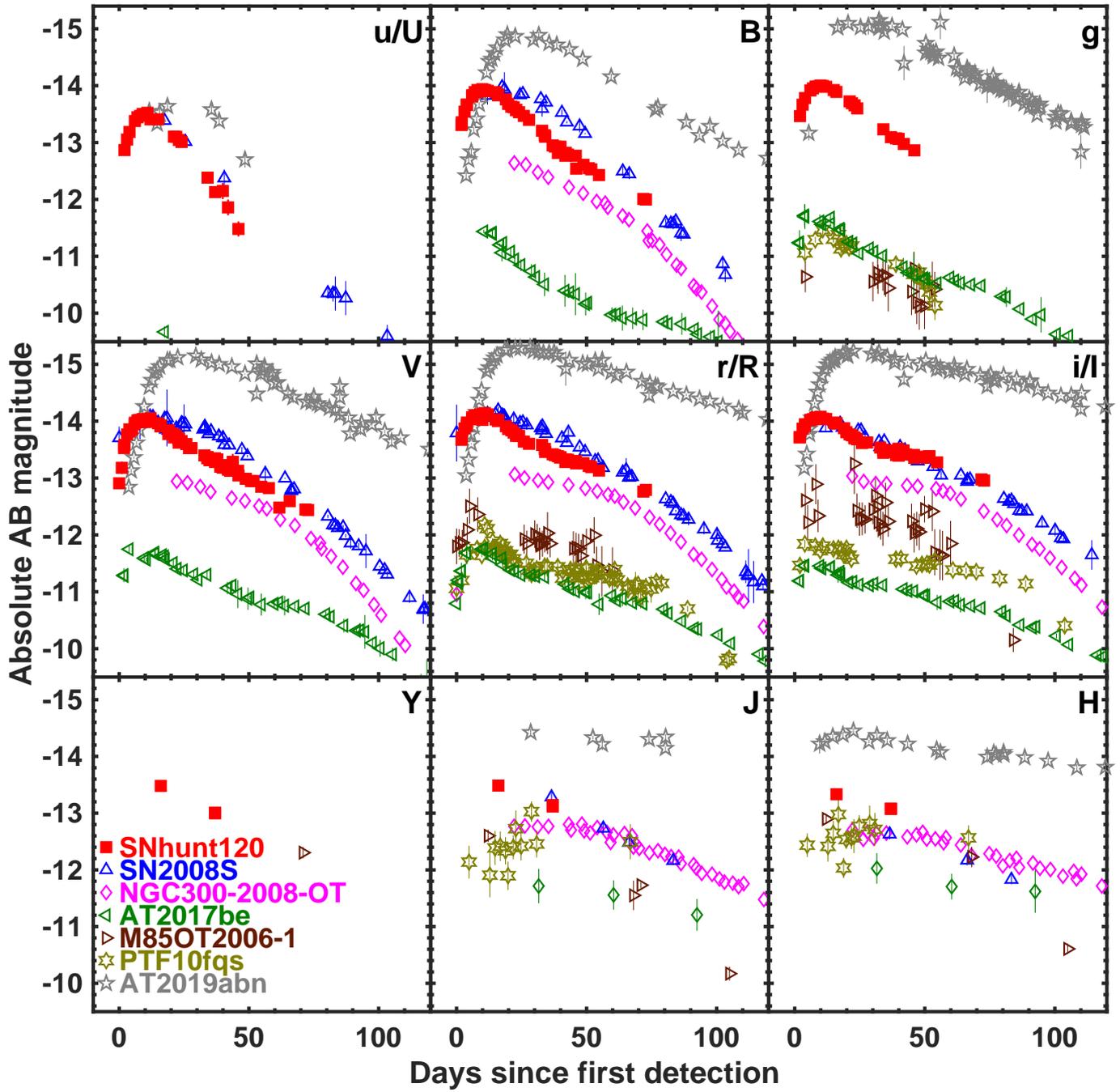}
\caption{Absolute magnitude light curves  shifted to the AB system of SNhunt120 compared to the ILRTs  M85-2006-OT1, SN~2008S, NGC~300-2008-OT1, PTF10fqs, AT~2017be, and AT~2019abn.}  
\label{Fig:abmagSNhunt120}
\end{figure}

\clearpage
\begin{figure}
\centering
\includegraphics[width=18cm]{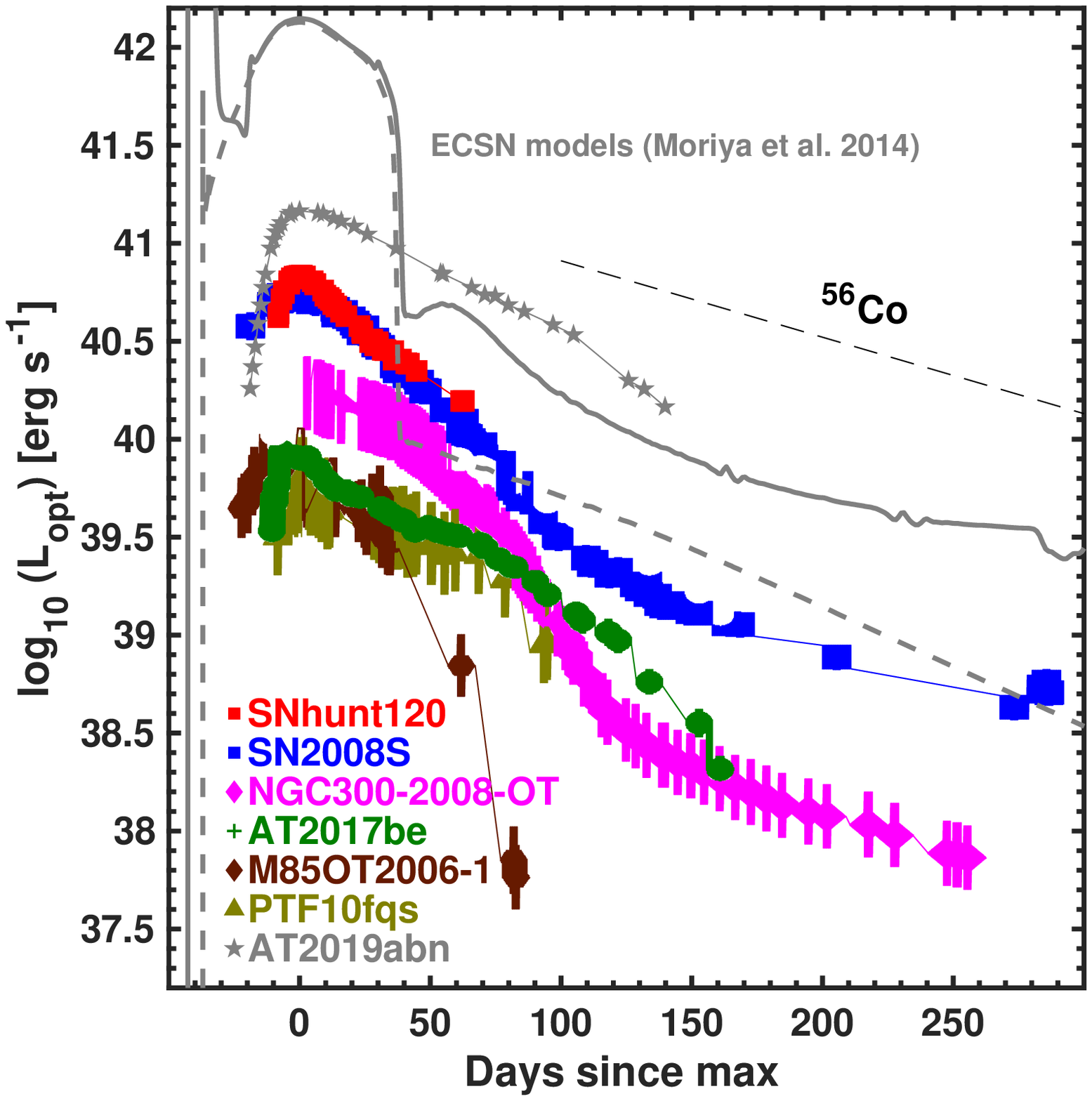}
\caption{Pseudo-bolometric light curve of SNhunt120 compared to the light curves of our EC SN comparison sample from \citet{at2017be} and AT~2019abn \citep{jencson2019,at2019abn}.  The light curves of two EC SN models from \citet{moriya14} without CSI  (dashed line) and a model with (solid line) CSI are also plotted. The dashed straight line corresponds to the complete trapping of the  $^{56}$Co $\rightarrow$ $^{56}$Fe energy deposition.}
\label{Fig:bolocompSNhunt120}
\end{figure}

%Added by TeX Support
\clearpage
\begin{deluxetable}{ccccccccc}
\tablecaption{Photometry of the local sequence for SNhunt120 in the `standard' system.\label{tab:SNhunt120_opt_locseq}\tablenotemark{a}}
   \tablehead{
      \colhead{ID} & \colhead{$\alpha$ (2000)} & \colhead{$\delta$ (2000)} & 
      \colhead{$B$} & \colhead{$V$} & \colhead{$u^\prime$} & \colhead{$g^\prime$} & 
      \colhead{$r^\prime$} & \colhead{$i^\prime$} }
   \startdata
   1 & 223.501175 &   3.554772 & \ldots      & \ldots      & \ldots      & \ldots      & \ldots      & \ldots     \\
  2 & 223.494568 &   3.512511 & \ldots      & \ldots      & \ldots      & \ldots      & \ldots      & \ldots     \\
  3 & 223.450409 &   3.549971 & \ldots      & \ldots      & \ldots      & \ldots      & \ldots      & \ldots     \\
  4 & 223.491287 &   3.578468 & 16.572(028) & 16.018(011) & 17.253(048) & 16.239(009) & 15.847(019) & 15.692(010)\\
  5 & 223.454926 &   3.543110 & 16.771(020) & 16.082(012) & 17.797(071) & 16.387(010) & 15.892(007) & 15.716(010)\\
  6 & 223.473694 &   3.598971 & 17.385(050) & 16.734(025) & 18.245(055) & 17.016(035) & 16.567(029) & 16.392(021)\\
  7 & 223.545074 &   3.513310 & 17.712(029) & 17.034(024) & 18.675(033) & 17.374(035) & 16.854(013) & 16.661(024)\\
  8 & 223.423279 &   3.524919 & 18.486(023) & 17.988(049) & 19.094(043) & 18.198(036) & 17.827(034) & 17.709(060)\\
  9 & 223.409164 &   3.544031 & 18.709(027) & 18.131(044) & 19.455(207) & 18.315(119) & 17.930(072) & 17.795(068)\\
 10 & 223.525085 &   3.558814 & 19.137(166) & 18.274(018) & 20.034(082) & 18.703(032) & 17.994(079) & 17.723(096)\\
 11 & 223.461456 &   3.557315 & 19.394(061) & 17.967(060) & 19.917(172) & 18.701(003) & 17.397(041) & 16.743(016)\\
 12 & 223.466751 &   3.512445 & 19.219(068) & 18.595(080) & 19.980(066) & 18.800(085) & 18.276(096) & 18.154(070)\\
 13 & 223.450058 &   3.525849 & 19.625(047) & 18.460(053) & \ldots      & 18.985(028) & 17.875(043) & 17.522(032)\\
 14 & 223.413147 &   3.509561 & 19.529(075) & 18.903(025) & 19.950(072) & 19.241(025) & 18.724(024) & 18.467(052)\\
 15 & 223.500031 &   3.629244 & 19.735(062) & 18.549(084) & \ldots      & 19.149(033) & 17.913(043) & 17.394(027)\\
 16 & 223.488373 &   3.505094 & 19.659(061) & 18.660(035) & \ldots      & 19.168(068) & 18.195(060) & 17.927(074)\\
 17 & 223.537628 &   3.636855 & 19.683(100) & 19.045(044) & 20.249(094) & 19.355(005) & 18.847(103) & 18.771(035)\\
 18 & 223.432632 &   3.621403 & 19.990(188) & 19.316(039) & 20.400(132) & 19.637(111) & 19.189(035) & 18.764(062)\\
 19 & 223.469666 &   3.623828 & 19.806(138) & 19.236(086) & \ldots      & 19.574(057) & 18.878(003) & 18.537(061)\\
 20 & 223.407272 &   3.620129 & 19.984(091) & 19.721(097) & 19.100(063) & 20.079(100) & 19.521(052) & 19.187(014)\\
 21 & 223.508453 &   3.619909 & 20.646(152) & 19.390(189) & \ldots      & 20.051(112) & 18.770(084) & 18.039(116)\\
\enddata  
\tablenotetext{a}{Note. -- Values in parenthesis are 1-$\sigma$ uncertainties that correspond to the rms of the instrumental errors of the photometry obtained over a minimum of three nights observed relative to standard star fields.}
\end{deluxetable}

\begin{deluxetable}{ccccccccc}
%\rotate
%\tabletypesize{\scriptsize}
%\tablecolumns{9}
\tablewidth{0pt}
\tablecaption{NIR photometry of the local sequences for SNhunt120 in the `standard' system.\tablenotemark{a}\label{tab:SNhunt120_nir_locseq}}
\tablehead{
\colhead{ID} &
\colhead{$\alpha (2000)$} &
\colhead{$\delta (2000)$} &
\colhead{$Y$} &
\colhead{N}   & 
\colhead{$J_{rc2}$} &
\colhead{N}   & 
\colhead{$H$} &
\colhead{N}   }
\startdata
101 & 223.4504 & 3.549965 & 13.60(01) & 4 & 13.40(34) & 5	& 12.73(03) & 5 \\   
102 & 223.4811 & 3.582550 & 13.65(02) & 5 & 13.41(03) & 6	& 13.07(02) & 6 \\
103 & 223.4913 & 3.578437 & 15.10(02) & 5 & 14.84(03) & 6	& 14.47(03) & 6 \\
104 & 223.4615 & 3.557310 & 15.73(04) & 5 & 15.25(04) & 6	& 14.57(04) & 6 \\
105 & 223.4739 & 3.598759 & 15.66(04) & 1 & 15.52(03) & 2	& 15.23(06) & 1 \\
106 & 223.4524 & 3.570402 & 16.97(10) & 4 & 16.37(15) & 5	& 15.76(10) & 5 \\
107 & 223.4747 & 3.582985 & 17.47(21) & 3 & 17.13(20) & 3	& 16.53(19) & 3 \\
\enddata
\tablenotetext{a}{Note. -- Values in parenthesis are 1-$\sigma$ uncertainties and correspond to an rms of the instrumental errors of the photometry obtained
over the number of  photometric nights N that standard star fields were observed.}
\end{deluxetable}

\clearpage
\begin{deluxetable}{cc|cc|cc|cc|cc|cc}
\rotate
\tabletypesize{\scriptsize}
\tablewidth{0pt}
%\tablecolumns{18}
\tablecaption{Optical photometry of SNhunt120 in the Swope `natural' system.\tablenotemark{a}\label{tab:SNhunt120_optphot}}
\tablehead{
\colhead{Time\tablenotemark{\dag}}&
\colhead{$u$}&

\colhead{Time\tablenotemark{\dag}}&
\colhead{$B$}&

\colhead{Time\tablenotemark{\dag}}&
\colhead{$g$}&

\colhead{Time\tablenotemark{\dag}}&
\colhead{$V$}&

\colhead{Time\tablenotemark{\dag}}&
\colhead{$r$}&

\colhead{Time\tablenotemark{\dag}}&
\colhead{$i$} }
\startdata
15.82 & 19.386(0.026) &  15.80 & 18.823(0.018) & 15.82 & 18.512(0.015) & 15.79 & 18.247(0.017) & 15.81 & 17.990(0.014) & 15.81 & 17.814(0.016) \\
16.84 & 19.201(0.024) &  16.86 & 18.637(0.014) & 16.84 & 18.311(0.011) & 16.87 & 18.062(0.013) & 16.85 & 17.835(0.011) & 16.85 & 17.665(0.013) \\
17.77 & 19.068(0.025) &  17.75 & 18.519(0.016) & 17.78 & 18.195(0.013) & 17.74 & 17.972(0.016) & 17.76 & 17.731(0.011) & 17.76 & 17.587(0.017) \\
19.84 & 18.875(0.022) &  19.85 & 18.345(0.013) & 19.83 & 18.060(0.009) & 19.86 & 17.855(0.011) & 19.83 & 17.637(0.009) & 19.84 & 17.510(0.009) \\
20.82 & 18.810(0.104) &  20.80 & 18.293(0.013) & 20.82 & 18.025(0.009) & 20.81 & 17.820(0.010) & 20.83 & 17.605(0.011) & 20.84 & 17.496(0.010) \\
21.88 & 18.786(0.020) &  21.84 & 18.256(0.010) & 21.86 & 17.993(0.008) & 21.85 & 17.808(0.010) & 21.86 & 17.605(0.009) & 21.87 & 17.468(0.009) \\
22.86 & 18.752(0.032) &  22.84 & 18.279(0.016) & 22.87 & 17.986(0.010) & 22.83 & 17.792(0.011) & 22.86 & 17.590(0.009) & 22.85 & 17.476(0.009) \\
23.77 & 18.773(0.036) &  23.75 & 18.273(0.020) & 23.77 & 17.973(0.013) & 23.74 & 17.794(0.015) & 23.76 & 17.577(0.014) & 23.76 & 17.455(0.015) \\
24.68 & 18.730(0.057) &  24.69 & 18.255(0.027) & 24.67 & 17.983(0.018) & 24.70 & 17.800(0.018) & 24.68 & 17.564(0.013) & 24.69 & 17.490(0.013) \\
25.85 & 18.846(0.069) &  25.83 & 18.274(0.026) & 25.85 & 17.997(0.019) & 25.82 & 17.774(0.017) & 25.84 & 17.589(0.012) & 25.84 & 17.493(0.012) \\
28.83 & 18.846(0.031) &  28.80 & 18.293(0.013) & 28.82 & 18.054(0.011) & 28.81 & 17.865(0.013) & 28.82 & 17.678(0.011) & 28.83 & 17.557(0.009) \\
$\cdots$& $\cdots$    &  29.80 & 18.349(0.014) & 29.81 & 18.086(0.010) & 29.81 & 17.904(0.010) & 29.83 & 17.686(0.009) & 29.84 & 17.567(0.010) \\
34.89 & 19.151(0.030) &  34.87 & 18.549(0.015) & 34.90 & 18.253(0.010) & 34.87 & 18.057(0.011) & 34.89 & 17.868(0.011) & 34.88 & 17.766(0.011) \\
35.77 & 19.153(0.024) &  35.75 & 18.558(0.017) & 35.78 & 18.285(0.012) & 35.74 & 18.090(0.014) & 35.77 & 17.908(0.012) & 35.76 & 17.781(0.014) \\
36.79 & 19.209(0.028) &  36.77 & 18.603(0.014) & 36.80 & 18.323(0.011) & 36.77 & 18.138(0.012) & 36.79 & 17.950(0.010) & 36.78 & 17.834(0.014) \\
37.84 & 19.244(0.026) &  37.82 & 18.646(0.014) & 37.84 & 18.377(0.010) & 37.81 & 18.179(0.012) & 37.83 & 17.967(0.010) & 37.82 & 17.852(0.011) \\
47.87 & 19.871(0.067) &  47.85 & 19.078(0.019) & 47.88 & 18.745(0.014) & 47.84 & 18.459(0.013) & 47.86 & 18.248(0.012) & 47.86 & 18.089(0.014) \\
50.78 & 20.125(0.086) &  50.76 & 19.230(0.030) & 50.79 & 18.882(0.020) & 50.75 & 18.546(0.021) & 50.77 & 18.289(0.012) & 50.77 & 18.076(0.015) \\
$\cdots$ & $\cdots$   &  52.75 & 19.363(0.042) & 52.77 & 18.904(0.023) & 52.76 & 18.629(0.025) & 52.78 & 18.323(0.015) & 52.79 & 18.129(0.015) \\
53.74 & 20.105(0.142) &  53.70 & 19.264(0.045) & 53.72 & 18.912(0.027) & 53.71 & 18.646(0.026) & 53.73 & 18.343(0.015) & 53.73 & 18.140(0.016) \\
55.76 & 20.394(0.139) &  55.78 & 19.360(0.059) & 55.74 & 19.002(0.030) & 55.77 & 18.682(0.030) & 55.75 & 18.379(0.022) & 55.73 & 18.115(0.026) \\
59.77 & 20.773(0.133) &  59.79 & 19.638(0.039) & 59.78 & 19.114(0.016) & 59.79 & 18.782(0.019) & 59.78 & 18.449(0.013) & 59.76 & 18.179(0.014) \\
\enddata
\tablenotetext{a}{Note. -- Values in parenthesis are 1-$\sigma$ uncertainties  corresponding to the sum in quadrature of the instrumental error and the nightly zero-point error.}
\tablenotetext{\dag}{Note. -- JD+2456000.}
\end{deluxetable}

\begin{deluxetable}{c c c c}
%\tablecolumns{4}
\tablewidth{0pt}
\tablecaption{NIR photometric observations of SNhunt120 in the du Pont (+ RetroCam) `natural' system.\tablenotemark{a}  \label{tab:SNhunt-nirphot}}
\tablehead{
\colhead{JD}  & \colhead{Photometry [mag]} & \colhead{error [mag]} & \colhead{filter [mag]}}
\startdata
2456029.82 & 17.21 & 0.02 &      $Y$       \\         
2456050.74 & 17.68 & 0.01 &      $Y$       \\
2456050.75 & 17.70 & 0.01 &      $Y$       \\
2456029.87 & 16.85 & 0.02 &   $J_{rc2}$       \\
2456050.78 & 17.22 & 0.02 &   $J_{rc2}$       \\
2456050.79 & 17.20 & 0.02 &   $J_{rc2}$       \\
%2456169.50 & 18.151 & 0.055 &   $Jrc2?$ S/N=3\\
2456029.85 & 16.50 & 0.02 &   $H$   \\  
2456050.77 & 16.75 & 0.03 &   $H$  \\
2456050.77 & 16.76 & 0.03 &    $H$  \\
%2456168.52 & 17.921 & 0.090 &    $H$ S/N=2.3\\
\enddata
\tablenotetext{a}{Note. -- The uncertainties in photometry correspond to the sum in quadrature of the instrumental error and the nightly zero-point error.}
\end{deluxetable}

\clearpage 
\begin{deluxetable}{ccc}
\tablewidth{0pt}
\tablecaption{LSQ $gr$-band photometry of SNhunt120 in the Swope `natural' system.\label{tab:SNhunt120_LSQ_phot}}
\tablehead{
\colhead{JD}&
\colhead{$V^{a}$}&
\colhead{err}}
\startdata
2456013.80&  18.951  & 0.054  \\  
2456014.80&  18.678  & 0.108  \\ 
2456017.79&  18.042  & 0.010  \\ 
2456018.79&  18.001  & 0.060  \\ 
2456021.78&  17.873  & 0.017  \\ 
2456022.78&  17.858  & 0.019  \\ 
2456024.70&  17.850  & 0.100  \\ 
2456032.73&  18.085  & 0.012  \\ 
2456034.85&  18.144  & 0.013  \\ 
2456036.71&  18.237  & 0.036  \\ 
2456040.70&  18.332  & 0.019  \\ 
2456048.62&  18.545  & 0.022  \\ 
2456050.66&  18.511  & 0.057  \\ 
2456057.57&  18.573  & 0.140  \\ 
2456071.50&  19.035  & 0.049  \\ 
2456075.53&  19.377  & 0.082  \\ 
2456079.51&  19.264  & 0.079  \\
\enddata
\tablenotetext{a}{Note. -- Photometry is computed relative to the $V$-band magnitude  of star with identification (ID) number 3 in Table~\ref{tab:SNhunt120_opt_locseq}.  To facilities good agreement between the LSQ $gr$-band and CSP-II $V$-band light curves, a 0.03 mag offset was added to the CSP-II magnitude of star 3 as explained in the text.}
\end{deluxetable}

\begin{deluxetable}{cccc}
%\floattable
%\rotate
%\tabletypesize{\scriptsize}
%\tablecolumns{}
\tablewidth{0pt}
\tablecaption{Spitzer fluxes for SNhunt120.\label{tab:spitzer}}
\tablehead{
\colhead{MJD} &
\colhead{3.6 $\mu$m} &
\colhead{4.5 $\mu$m} \\
\colhead{} &
\colhead{} &
\colhead{}}
\startdata
56041.24  & 5.60$\times10^{-18}\pm2.66\times10^{-20}$ & 3.61$\times10^{-18}\pm2.19\times10^{-20}$\\
56052.83  & 4.95$\times10^{-18}\pm2.68\times10^{-20}$ & 3.30$\times10^{-18}\pm2.08\times10^{-20}$\\
57168.52  & 8.90$\times10^{-19}\pm1.68\times10^{-19}$     & 1.18$\times10^{-18}\pm7.73\times10^{-21}$\\
\enddata
\tablenotetext{a}{Note. -- fluxes in erg s$^{-1}$ \AA$^{-1}$ cm$^{-2}$.}
\end{deluxetable}

\clearpage
\clearpage 
\begin{deluxetable}{l l l c l l c}
\tablewidth{0pt}
\tablecolumns{7}
\tablecaption{\label{tab:specobs} Journal of Spectroscopic Observations of SNhunt120.}
\tablehead{
\colhead{Object} & 
\colhead{Date} & 
\colhead{Date} &	
\colhead{Days since} & 
\colhead{Telescope} & 
\colhead{Instrument}  & 
\colhead{Resolution}\\
\colhead{} & 
\colhead{(JD)} &
\colhead{(UT)} & 
\colhead{discovery\tablenotemark{a}} &
\colhead{} &
\colhead{}   & 
\colhead{(\AA)}}
\startdata
SNhunt120 & 2456014.66 & March 28.16     & $+$0.7  & du Pont & WFCCD         & 7.5\\
SNhunt120 & 2456025.79 & April 08.29     & $+$11.8 & Baade   & FIRE          & 22 \\
SNhunt120 & 2456049.65 & May 02.15       & $+$35.7 & du Pont & WFCCD         &  7.5\\
SNhunt120 & 2456054.75 & May 07.25       & $+$40.8 & Baade   & FIRE          & 22\\
\enddata
\tablenotetext{a}{Days since outburst assuming outburst date  of 25.15 March 2012 UT (JD$-$2456011.7.}
\end{deluxetable}

\begin{deluxetable}{l c c c c }
\tablewidth{0pt}
%\tablecolumns{5}
\tablecaption{\label{tab:SNhunt120peaks} Phase and value of peak absolute magnitudes of SNhunt120.}
\tablehead{
\colhead{Band $X$} & 
\colhead{Phase$^*$} & 
	\colhead{$m_{X}$} &	
\colhead{$M_X$} & 
\colhead{Reddening corrected $M_X$} }
\startdata
$u$ & 9.72$\pm$0.4  & 18.77$\pm$0.01 & $-$13.58$\pm$0.14 & $-$13.43$\pm$0.29   \\
$B$ &10.00$\pm$0.2  & 18.25$\pm$0.01 & $-$14.09$\pm$0.14 & $-$13.80$\pm$0.29   \\
$g$ & 9.62$\pm$0.2  & 17.98$\pm$0.01 & $-$14.37$\pm$0.14 & $-$13.98$\pm$0.29   \\
$V$ & 9.59$\pm$0.2  & 17.78$\pm$0.01 & $-$14.56$\pm$0.14 & $-$14.03$\pm$0.29   \\
$r$ & 9.41$\pm$0.2  & 17.57$\pm$0.01 & $-$14.77$\pm$0.14 & $-$14.12$\pm$0.29   \\
$i$ & 9.11$\pm$0.2  & 17.46$\pm$0.01 & $-$14.88$\pm$0.14 & $-$14.07$\pm$0.29   \\
\enddata
\tablenotetext{*}{Note. -- Days since first detection.}
\end{deluxetable}

%In [8]: print s.u.Tmax,s.u.e_Tmax,s.u.Mmax,s.u.e_Mmax,s.u.dm15,s.u.e_dm15
%56023.7881886 0.366641777715 18.7417057568 0.0135115012431 0.592296018397 0.0392074460951

%In [10]: print s.g.Tmax,s.g.e_Tmax,s.g.Mmax,s.g.e_Mmax,s.g.dm15,s.g.e_dm15
%56023.2707476 0.186665371382 17.9705847182 0.00467379313741 0.439623263307 0.0134472917582

%In [12]: print s.r.Tmax,s.r.e_Tmax,s.r.Mmax,s.r.e_Mmax,s.r.dm15,s.r.e_dm15
%56023.21774 0.193182328854 17.5739046232 0.0052150568729 0.43415767381 0.0119103559822

%In [14]: print s.i.Tmax,s.i.e_Tmax,s.i.Mmax,s.i.e_Mmax,s.i.dm15,s.i.e_dm15
%56023.014407 0.191932536079 17.4646509534 0.00501273636902 0.417939792276 0.0119443096808

%print s.B.Tmax,s.B.e_Tmax,s.B.Mmax,s.B.e_Mmax,s.B.dm15,s.B.e_dm15
%56023.7227636 0.242541037107 18.2331944566 0.00830497679003 0.47415887525 0.0181879235405

%print s.V.Tmax,s.V.e_Tmax,s.V.Mmax,s.V.e_Mmax,s.V.dm15,s.V.e_dm15
%56023.3635185 0.198802545918 17.779732793 0.00599344650687 0.42677462764 0.0150545009402

%example of how I com[uted the absolute magnitudes
%In [74]: print 18.23- (AVtot*1.321)   -  32.3        
% print math.sqrt(0.15**2 + 0.25**2 + 0.01**2)

\begin{deluxetable}{lcc}
\tablewidth{0pt}
%\tablecolumns{5}
\tablecaption{\label{tab:comp_obj_snhunt120} Adopted color excess values and distance moduli for the ILRT sample compared to SNhunt120.}
\tablehead{
\colhead{ILRT} & 
\colhead{$\mu$ } & 
\colhead{$E(B-V)_{tot}$} \\
\colhead{} & 
\colhead{(mag) } & 
\colhead{(mag)}}
\startdata
%ILRT              & $\mu$  &$E(B-V)_{tot}$\\
M85-OT2006-1      & 31.03 & 0.14 \\
SN~2008S          & 31.34 & 0.69 \\
NGC~300-2008-OT1  & 28.78 & 0.40 \\
PTF10fqs          & 26.37 & 0.04\\
AT~2017be         & 31.15 & 0.09 \\
%AT~2018hso        & 29.47 & 0.30 \\
AT~2019abn        & 31.64 & 0.85 \\
\enddata
%\tablenotetext{*}{Note. -- Days since first detection.}
\end{deluxetable}

%\clearpage 
\begin{deluxetable}{lcccccc}
%\tablecolumns{7}
\tablewidth{0pt}
\tablecaption{\label{tab:FWHM_SNhunt120} FWHM velocity measurements of prominent spectral features in SNhunt120.}
\tablehead{
\colhead{Phase$^{a}$} &
\colhead{H$\beta$} &
\colhead{H$\alpha$} &
%\colhead{[\ion{Ca}{ii}] $\lambda$7291} &
%\colhead{[\ion{Ca}{ii}] $\lambda$7324} &
\colhead{\ion{Ca}{ii} $\lambda$8662} &
\colhead{\ion{He}{i} $\lambda$10830} &
\colhead{P$\gamma$} &
\colhead{P$\beta$} \\
\colhead{(days)} &
\colhead{(km s$^{-1}$)} &
\colhead{(km s$^{-1}$)} &
\colhead{(km s$^{-1}$)} &
%\colhead{(km s$^{-1}$)} &
%\colhead{(km s$^{-1}$)} &
\colhead{(km s$^{-1}$)} &
\colhead{(km s$^{-1}$)} &
\colhead{(km s$^{-1}$) }}
\startdata
%+1  & 1413$\pm$122 & $739\pm21$ & $954\pm97$  &$\cdots$ & $\cdots$ & $\cdots$  \\
%+12   & $\cdots$ & $\cdots$ & $\cdots$  &$\cdots$ & $1326\pm196$ & $1500\pm277$  \\

+1  & 1362$\pm$122 & 636$\pm$21 & $\cdots$  &$\cdots$ & $\cdots$ & $\cdots$  \\

+12   & $\cdots$ & $\cdots$ & 742$\pm$97  &$\cdots$ & $1182\pm196$ & 1375$\pm$277  \\

%+36 & $785\pm$115  & $584\pm17$ &  $671\pm113$ &$1776\pm831$ & $\cdots$     & $\cdots$ \\
%+41 & $\cdots$  & $\cdots$ &  $\cdots$ & $\cdots$ & $\cdots$     & $1918\pm690$ \\

+36 & 690$\pm$115  & 448$\pm$17 &  $\cdots$ $\cdots$ & $\cdots$     & $\cdots$ \\

+41 & $\cdots$  & $\cdots$ & 301$\pm$113  &$1672\pm831$ & $\cdots$     & $1822\pm690$ \\
\enddata
\tablenotetext{a}{Note. -- Phase with respect to date of discovery.}
\end{deluxetable}

\clearpage 

\begin{appendix}

\section{Host metallicity}
\label{appendixA}

\begin{figure}[!htb]
\centering
\includegraphics[width=12cm]{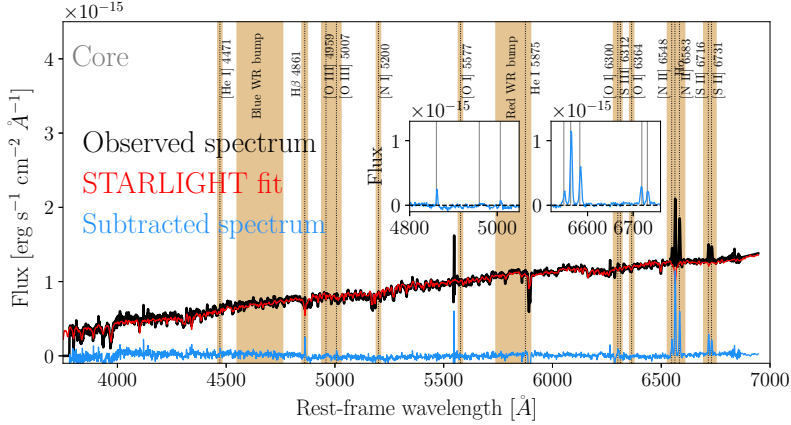}
\caption{Visual wavelength spectrum of the nucleus of NGC~5775 shown in the host-galaxy rest frame (in black), and the resulting single stellar population synthesis computed by STARLIGHT (in red). The continuum-subtracted spectrum is shown in blue, where the most prominent host nebular emission features are labeled. The nebular lines  shown within the insets provide flux ratios indicating a  metallicity higher than solar of $8.7\pm0.2$ dex on the O3N2 scale at the core of the host. 
When we adopt the metallicity gradient from \citet{pilyugin04} of $-$0.47 dex~R$_{25}^{-1}$ , the metallicity at the position of SNhunt120 is estimated to be 8.34 dex. When we assume an uncertainty of 0.1 dex~R$_{25}^{-1}$ for the gradient, the uncertainty on the metallicity at the SN position would be 0.22 dex.}
\label{fig:SDSShost}
\end{figure}

We plot in Fig.~\ref{fig:SDSShost} the SDSS spectrum of NGC~5775 and also indicate and label nebular emission lines. In particular, emission features associated with 
H$\alpha$, [\ion{N}{ii}] $\lambda$6584, and [\ion{S}{ii}]  $\lambda\lambda$6717,6731   are labeled and fit with a single-Gaussian function in order to obtain  a measure of the observed wavelength and line flux.
From these features we compute an average redshift of $z=0.00583\pm0.00005$.

The spectrum  provides a means for estimating an oxygen metallicity abundance, but with respect to its central region. To do so, the SDSS spectrum was first fit with a modified version of {\sc STARLIGHT} (\citealt{2005MNRAS.358..363C,2016MNRAS.458..184L}, priv. comm.), a program that models the stellar component of the spectral continuum by estimating the fractional contribution of simple stellar populations (SSP) of different ages and metallicities, adding a dust attenuation foreground screen.
The best-fit SSP model is then removed from the spectrum, and then  Gaussian functions are fit to [\ion{N}{ii}] and [\ion{O}{iii}] nebular lines. From these fits we obtain a metallicity higher than solar of 
$8.73\pm0.20$ dex on the O3N2 scale \citep{2004MNRAS.348L..59P}.

As this is an indication for the center region of the host and not for the position of SNhunt120, we extrapolated the galaxy nucleus value to the approximate location of SNhunt120 using the metallicity gradient from \citet{pilyugin04}  of $-$0.47 dex~R$_{25}^{-1}$.
 Adopting  values for the diameter and position angle for NGC~5775 from  NED and Hyperleda \citep{makarov2014},\footnote{Hyperleda is a database of various physical properties of galaxies and can be found at \url{http://leda.univ-lyon1.fr}} we compute 
a  radius-normalized distance from the host nucleus of 0.82. Altogether, we find a  local oxygen abundance for SNhunt120 of 12~+~log(O/H)~ $= 8.34\pm0.22$ dex, where we adopt a 0.1 dex~R$_{25}^{-1}$ uncertainty on the gradient.

\clearpage 
\section{Galactic and host reddening}
\label{appendixB}

NED lists the Galactic reddening along the line of sight to NGC~5775 as inferred by  the \citet{2011ApJ...737..103S} (SF11) recalibration of the \citet{1998ApJ...500..525S} infrared-based dust map to be  $E(B-V)_{MW} = 0.037$ mag. When a standard $R^{MW}_V = 3.1$ value is adopted, this gives a foreground visual extinction of $A^{MW}_{V} = 0.115$ mag.

When estimating the host reddening of SNhunt120, we note that there are no reliable methods to do so, and regardless of the adopted method, it is accompanied by significant uncertainty. With this caveat in mind, inspection of the two low-resolution visual wavelength spectra presented below reveals a \ion{Na}{i}~D absorption feature located at the expected location given the redshift of the host galaxy.
In principle, a reliable measure of column density can provide a measure of  the host extinction \citep[e.g.,][see their Fig.~9]{2013ApJ...779...38P}.
Unfortunately, we are
only able to estimate rough column densities
from these low-dispersion spectra as they do not resolve the individual components that contribute to the absorption.
Measuring the EW of the \ion{Na}{i} feature from the low-resolution spectra indicates a value of  EW$_{\rm \ion{Na}{i}~D} \approx 2.5\pm0.5$~\AA. This   value  lies outside the  range of validity for the \citet{2013ApJ...779...38P} relation.  
 We next searched for evidence of a diffuse interstellar band (DIB) located at 5780~\AA\ that  is known to be correlated with  $A^{host}_V$ \citep{2013ApJ...779...38P}. 
Unfortunately, the signal-to-noise ratio of our spectra is   too low to provide any meaningful measurement of this DIB feature. 

To approximately estimate the reddening of SNhunt120, we compared its observed broadband colors to the intrinsic broadband colors of other ILRTs.
Fig.~\ref{Fig:colorSNhunt120} shows that the  $(V-r)$ and $(V-R)$ color evolution of the comparison sample  is rather similar  to that of SNhunt120. Here the broadband colors of the comparison objects were corrected for  reddening using values found in the literature. 
When we assume that the intrinsic colors of SNhunt120 are similar to those of the bluest objects in the comparison sample,  the host-galaxy color excess affecting SNhunt120 can be estimated by shifting its observed broadband colors to  match the intrinsic color curves of the bluest  objects in the comparison sample. 
In doing so, we find that the observed colors agree with the intrinsic colors of the bluest objects with $E(V-r)_{host} = 0.095\pm0.037$ mag. When a \citet{NED_extinction} reddening law is adopted, this translates into  a host visual extinction  of  $A^{host}_{V} = 0.64\pm0.25$~mag. Values exceeding those contained within the quoted uncertainty lead to poor agreement between SNhunt120 and the bluest objects in Fig~\ref{Fig:colorSNhunt120}.  When $A^{host}_{V}$ is combined with $A^{MW}_{V}$ , we obtain a total  visual extinction   of $A^{tot}_{V} = 0.76\pm0.25$ mag for SNhunt120. 

Objects such as SNhunt120 are typically enshrouded by circumstellar dust, and as a result,  the  dust may follow a peculiar reddening law. When we consider the relations for reddening assuming graphitic (with grains of size 0.2~$\mu$m) circumstellar  dust as  presented by \citet{2012ApJ...759...20K}, $A_{\lambda}$ would be 17\% lower at most than that from the \citet{NED_extinction} law in the optical ($i$ band). If, on the other hand, the dust is formed by silicates (with grains of size 0.2~$\mu$m), $A_{\lambda}$ would be 45\% lower at most than the \citet{NED_extinction} law in the optical ($i$ band), and up to 30\% higher in the $u$ band.

\end{appendix}
\end{document}